\documentclass[12pt]{article}
\usepackage{amsmath}
\usepackage{tensor}
\usepackage{amssymb}
\usepackage{mathrsfs}
\usepackage{hyperref}
\usepackage{slashed}
\usepackage{enumerate}
\usepackage{cite}
\numberwithin{equation}{section}
\title{Double Field Theory and $\alpha'$ corrections:\\ explicit examples}
\author{Oleg Lunin\footnote{olunin@albany.edu}\ \ and Parita Shah\footnote{pshah3@albany.edu}}
\date{}
\begin{document}
\def\be{\begin{equation}}
\def\bea{\begin{eqnarray}}
\def\ee{\end{equation}}
\def\eea{\end{eqnarray}}
\def\d{\partial}
\def\eps{\varepsilon}
\def\la{\lambda}
\def\b{\bigskip}
\def\nn{\nonumber \\}
\def\p{\partial}
\def\t{\tilde}
\def\h{{1\over 2}}
\def\be{\begin{equation}}
\def\bea{\begin{eqnarray}}
\def\ee{\end{equation}}
\def\eea{\end{eqnarray}}
\def\b{\bigskip}
\def\u{\uparrow}
\newcommand{\comment}[2]{#2}
\maketitle

\begin{center}
\ \vskip -1.2cm
{\em  Department of Physics,\\
University at Albany (SUNY),\\
1400 Washington Avenue,\\
Albany, NY 12222, USA
 }
 \end{center}

\vskip 0.08cm
\begin{abstract}
We construct several solutions of effective actions for string theories beyond the supergravity approximation utilizing the framework of the Double Field Theory (DFT). The DFT effective actions, which are well suited for accommodating abelian and non--abelian T dualities, are naturally written in terms of modified connections and curvatures with torsion.  We construct the most general solutions with vanishing modified torsion, and such geometries don't receive quantum corrections. We also compute the leading $\alpha'$ corrections to several supergravity solutions, including a pair of geometries related by a non--abelian T duality.
\end{abstract}
\newpage
\tableofcontents
\def\be{\begin{equation}}
\def\bea{\begin{eqnarray}}
\def\ee{\end{equation}}
\def\eea{\end{eqnarray}}
\def\d{\partial}
\def\eps{\varepsilon}
\def\la{\lambda}
\def\b{\bigskip}
\def\nn{\nonumber \\}
\def\p{\partial}
\def\t{\tilde}
\def\h{{1\over 2}}
\def\be{\begin{equation}}
\def\bea{\begin{eqnarray}}
\def\ee{\end{equation}}
\def\eea{\end{eqnarray}}
\def\b{\bigskip}
\def\u{\uparrow}
\newpage
\section{Introduction}

Dualities play a very important role in string theory, and over the last three decades they have been used to obtain many remarkable results, ranging from discovery of D--branes \cite{PolchW1,PolchW2,PolchW3,PolchW4,PolchW5} and M--theory \cite{MtheoryW1,MtheoryW2,MtheoryW3} to insights into physics of black holes \cite{BHtdualW1,BHtdualW2,BHtdualW3,BHtdualW4} and the 
AdS/CFT correspondence \cite{AdSCFTw1,AdSCFTw2,AdSCFTw3,AdSCFTw4}. In particular, in perturbative string theory, a prominent role is played by T dualities, which emerge upon compactifying systems on tori or other manifolds \cite{TdualW1,TdualW2,NATDw1,NATDw2,NATDw3,NATDw4,NATDw5,NATDw6}. Combinations of such dualities have been successfully used to generate geometries of charged black holes \cite{SenCvetW1,SenCvetW2,SenCvetW3,SenCvetW4,SenCvetW5} and gravity duals of various field theories \cite{TsTQFTw1,TsTQFTw2,TsTQFTw3,TsTQFTw4,TsTQFTw5,TsTQFTw6,TsTdef,BetaDefW1,
BetaDefW2,BetaDefW3,BetaDefW4,BetaDefW5}. From the worldsheet perspective, T duality can be nicely visualized by doubling the number of coordinates and imposing various restrictions \cite{TseytlDFT}. This approach culminated in the introduction of the Double Field Theory (DFT) in \cite{DFTw1,DFTw2,DFTw3,DFTw4,DFTw5,DFTw6,DFTw7}. Although DFT was originally formulated on the worldsheet, its most fruitful applications have been found at the level of effective actions on the target space, which were mostly analyzed in the supergravity approximation \cite{DFTclassW1,DFTclassW2,DFTclassW3,DFTclassW4,DFTclassW5,DFTclassW6,DFTclassW7,
DFTclassW8,
DFTclassW9,DFTclassW10,DFTclassW11,DFTclassW12,DFTclassW13}. The leading $\alpha'$-corrections were incorporated in the DFT formalism as well \cite{DFTalpW1,DFTalpW2,DFTalpW3,DFTalpW4,DFTalpW5,DFTalpW6,DFTalpW7,1507}, and the goal of our article is to explore such corrections for several classes of explicit examples. 
\bigskip

To perform a T duality, one gauges some symmetries of the $\sigma$-model and introduces Lagrange multipliers to ensure that the gauge field has no dynamical degrees of freedom. Then by integrating out either the Lagrange multipliers or the gauge fields, one recovers either the original theory or its dual. Implementation of this procedure gives rise to the Bucher's rules for the abelian duality \cite{OGBuscherW1,OGBuscherW2} and its extensions in the non--abelian case \cite{NATD101,NATDAdSCFTw1,NATDAdSCFTw2,NATDAdSCFTw3,
NATDAdSCFTw4,NATDAdSCFTw5,NATDAdSCFTw6,NATDAdSCFTw7,NATDAdSCFTw8,
NATDAdSCFTw9,NATDAdSCFTw10,NATDAdSCFTw11,NATDAdSCFTw12}. In the Double Field Theory one keeps coordinates on both the original manifold and its dual, then the metric, the inverse metric, and the NS--NS B field combine into a single matrix object, and various dualities can be viewed as linear transformations of this object. In particular, at the level of effective actions, the duality becomes more manifest if one uses connections and curvatures with torsion, which contain information about both the metric and the $B$ field. While at the supergravity level this has been well known \cite{DFTclassW1,DFTclassW2,DFTclassW3,DFTclassW4,DFTclassW5,DFTclassW6,DFTclassW7,
DFTclassW8,
DFTclassW9,DFTclassW10,DFTclassW11,DFTclassW12,DFTclassW13}, the expressions for $\alpha'$-corrections to the action in terms of the curvatures with torsion is a relatively recent development \cite{1507}. It is well known that the detailed structure of $\alpha'$ corrections depends on a prescription used in evaluation of string amplitudes 
\cite{BdR-1w1,BdR-1w2,MT,BdR=MT}, and a great advantage of the DFT prescription is that it preserves the symmetries associated with T duality. This implies that starting with any solution of the DFT action with $\alpha'$ corrections, one can generate new geometries by performing purely algebraic manipulations, without solving highly nonlinear differential equations.

At the level of supergravity, the procedure for generating new solutions via algebraic transformation has been extensively explored over the last three decades, and it has led to important insights into physics of black holes \cite{SenCvetW1,SenCvetW2,SenCvetW3,SenCvetW4,SenCvetW5} and gravity duals of various field theories \cite{TsTQFTw1,TsTQFTw2,TsTQFTw3,TsTQFTw4,TsTQFTw5,TsTQFTw6,TsTdef,BetaDefW1,
BetaDefW2,BetaDefW3,BetaDefW4,BetaDefW5}. These constructions involved reduction of string theory on tori, and therefore they relied on the abelian version of the T duality and DFT. Recently  these solution generating techniques were extended to the non--abelian case \cite{LuShah}, where algebraic transformations involved mixings between spheres rather than circles. In contrast to the abelian case, where the algebraic manipulations, often called TsT transformations
\cite{TsTold,TsTQFTw1,TsTQFTw2,TsTQFTw3,TsTQFTw4,TsTQFTw5,TsTQFTw6,TsTdef}, map exact solutions of string theory into exact ones, the non--abelian T duality leads to additional quantum corrections \cite{2007.07902,2007.07897,2007.09494}. Therefore, even starting with an exact string theory solution, such as flat space, and dualizing it along a sphere, one finds a geometry that acquires $\alpha'$ corrections. Furthermore, the starting geometry may not be exact as well, and the most famous example involves neutral black holes, which served as seed solutions for generating charged holes and branes \cite{SenCvetW1,SenCvetW2,SenCvetW3,SenCvetW4,SenCvetW5}. Exploration of $\alpha'$ corrections both before and after duality is the main goal of this article. 

In the DFT prescription, the quantum corrections to the supergravity action are naturally written not in terms of the curvature and the Kalb--Ramond field, but rather in terms of twisted curvatures $R^{(\pm)}_{\mu\nu\lambda\sigma}$ that contain both these ingredients \cite{1507}\footnote{The precise definitions of the twisted curvature and twisted spin connections $\omega_\mu^{(\pm)ab}$ are given in section \ref{SecDFRrev}.}. The quantum corrections can vanish when either  $R^{(\pm)}_{\mu\nu\lambda\sigma}=0$ or when there are cancellations between various terms. The first option is definitely easier, and interestingly it leads to a rather nontrivial result. While vanishing of the Riemann tensor implies that the space is flat, up to global identifications, we will show that, in physically interesting cases, vanishing of the twisted curvature implies that the geometry is a product of three--spheres and a flat space. Going beyond the spaces with vanishing twisted curvature, we explore the geometries that don't receive quantum corrections due to cancellations in the effective action and those that do. Finally, we apply abelian and non--abelian T dualities to some exact geometries and demonstrate that, as expected, quantum corrections are generated in the latter case, but not in the former. 

\bigskip

This paper has the following organization. In section \ref{SecDFRrev} we briefly review some known facts about $\alpha'$ corrections in the framework of the Double Field Theory. In particular, we observe that at the leading order, such corrections are naturally formulated in terms of twisted curvatures $R^{(\pm)}$, so in section \ref{SecRmin} we construct the most general background with vanishing $R^{(-)}$. Such geometries turn out to be products of group manifolds, which in physically interesting situations reduce to products of three--spheres and a flat space, and they have vanishing $R^{(+)}$ as well, so all leading corrections to the supergravity action disappear term--by--term. Spaces with vanishing twisted curvature admit a set of special frames, the analogs of standard frames in flat spaces, and  in section \ref{SecFrameSn} we extend the construction of such special frames to  spheres and hyperboloids of arbitrary dimensions. Going back to quantum corrections, in section \ref{SecNSsoln} we analyze some geometries with nontrivial $R^{(-)}$ which do receive quantum corrections as well as examples of solutions that don't due to nontrivial cancellations. In section \ref{SecGW}, we evaluate $\alpha'$ corrections for linear gravitational waves. Finally, in section \ref{SecTdual} we consider the action of T duality in the presence of $\alpha'$ corrections and discuss some examples exploiting both abelian and non--abelian versions of this duality. Some technical calculations supporting our conclusions are presented in appendices.

\section{Double Field Theory and the $\alpha'$ expansion}
\label{SecDFRrev}

Dualities play a very important role in string theory, and some of these symmetries become manifest in the formalism known as the Double Field Theory (DFT) \cite{TseytlDFT,DFTw1,DFTw2,DFTw3,DFTw4,DFTw5,DFTw6,DFTw7}, which has been extensively used to generate new supergravity solutions \cite{DFTclassW1,DFTclassW2,DFTclassW3,DFTclassW4,DFTclassW5,DFTclassW6,DFTclassW7,
DFTclassW8,
DFTclassW9,DFTclassW10,DFTclassW11,DFTclassW12,DFTclassW13}. In this formalism, one doubles the number of target space coordinates and introduces generalized frames and curvatures to make the $O(d,d)$ symmetry associated with T dualities manifest. In this section we briefly review some concepts associated with DFT focusing on its applications to string theory actions beyond the supergravity approximation. 

\bigskip

To define the Double Field Theory, one begins with introducing the doubled coordinates $X^M=(x^\mu,\tilde{x}_{\tilde{\mu}})$, where $x^\mu$ and $\tilde{x}_{\tilde{\mu}}$ represent the coordinates of the parent theory and of its dual. Index $\mu$ takes $d$ values, where $d$ is the dimension of the spacetime. The action is formulated in terms of the generalized metric $\mathcal{H}_{MN}$ on the $2d$--dimensional space, and to avoid introduction of non--physical degrees of freedom, one imposes the ``strong constraint'' \cite{DFTw1,DFTw2,DFTw3,DFTw4,DFTw5,DFTw6,DFTw7} by demanding trivial dependence on the dual coordinates. The DFT invariant actions have been constructed both for the supergravity approximation \cite{TseytlDFT,DFTw1,DFTw2,DFTw3,DFTw4,DFTw5,DFTw6,DFTw7} and for the leading $\alpha'$ corrections \cite{DFTalpW1,DFTalpW2,DFTalpW3,DFTalpW4,DFTalpW5,DFTalpW6,DFTalpW7}, and in the latter case one should carefully choose the computational scheme to ensure invariance under T dualities. The authors of \cite{1507} constructed a manifestly $O(d,d)$ invariant action whose reduction to component fields reproduced the $\alpha^\prime$-corrections in the Metsaev-Tseytlin \cite{MT} and Bergshoeff-de Roo \cite{BdR-1w1,BdR-1w2} schemes. This was done by introducing an $\alpha^\prime$-corrected deformation of the gauge transformations of the generalized frames. The DFT action obtained in \cite{1507} can be concisely written as
\bea\label{DFT action}
S_{\text{DFT}}=\int d^{2d}Xe^{-2\Phi}\left[\mathcal{R}+a\mathcal{R}^{(-)}+b\mathcal{R}^{(+)}+\mathcal{O}(\alpha^{\prime2})\right],
\eea
where $\mathcal{R}$ and $\mathcal{R}^{(\pm)}$ are generalized curvatures in the doubled space. The explicit expressions for these curvatures are rather complicated, and they can be found in \cite{1507}. The parameters $a$ and $b$ are proportional to $\alpha'$, and numerical coefficients depend on the type of the string theory. The relations between component form of (\ref{DFT action}) and the actions in the Metsaev-Tseytlin and the Bergshoeff-de Roo schemes were explored in \cite{1507, BdR=MT}. We will work in the generalized Bergshoeff-de Roo scheme. 

In the component form, the action (\ref{DFT action}) can be written as 
\begin{equation}\label{DFTActComp}
\begin{aligned}
S=\int dx\sqrt{-g}e^{-2\phi}\left[L^{(0)}+L^{(1)}+\mathcal{O}(\alpha^{\prime2})\right]\,, 
\end{aligned}
\end{equation}
where the superscript denotes the order of $\alpha^\prime$ and the zeroth order (two-derivative) Lagrangian for the NS--NS sector is given by the supergravity approximation:
\begin{equation}\label{L0}
L^{(0)}=R-4\nabla_\mu\phi\nabla^\mu\phi+4\nabla_\mu\nabla^\mu\phi-\frac{1}{12}H^2\,.   
\end{equation}
The four-derivative terms, i.e., the first order in $\alpha^\prime$-corrected Lagrangian, are given by \cite{1507, BdR-1w1,BdR-1w2}
\begin{equation}\label{L1}
\begin{aligned}
L^{(1)}&=\frac{(a-b)}{4}H^{\mu\nu\rho}\Omega_{\mu\nu\rho}+\frac{(a+b)}{8}H^{\mu\nu\rho}\partial_\mu(H_{\nu}{}^{ab}\omega_{\rho a b})\\
&-\frac{(a+b)}{8}\bigg[R_{\mu\nu\rho\sigma}R^{\mu\nu\rho\sigma}-\frac{3}{2}H^{\mu\nu\rho}H_{\mu\sigma\lambda}R_{\nu\rho}{}^{\sigma\lambda}+\frac{1}{24}H^{\mu\nu\rho}H_{\mu\sigma}{}^\lambda H_{\nu\lambda}{}^\delta H_{\rho\delta}{}^\sigma\\
&+\frac{1}{3}\nabla_\mu H_{\nu\rho\sigma}\nabla^\mu H^{\nu\rho\sigma}+\frac{1}{8}H_{\mu\rho\delta}H^{\mu\rho}{}_\lambda H_{\nu\sigma}{}^\delta H^{\nu\sigma\lambda}\bigg]\,.
\end{aligned}    
\end{equation}
The action (\ref{DFTActComp})--(\ref{L1}) is reproduced by imposing the strong constraint in the DFT--invariant expression (\ref{DFT action}) \cite{1507}. The result (\ref{DFTActComp})--(\ref{L1}) can be written in a more concise way by introducing modified field strength $\tilde{H}_{\mu\nu\rho}$ and Riemann tensors $R^{(\pm)}_{\mu\nu\rho\sigma}$ \cite{1507}\footnote{There has been some recent development in obtaining a natural form of Riemann tensor for the case of generalized geometry and it would be interesting to consider expressing the $\alpha^\prime$-corrections using this new fundamental object \cite{fundRiem}.},
\begin{equation}\label{rpm action}
\begin{aligned}
S=\int dx\sqrt{-g}e^{-2\phi}\bigg(&R-4\nabla_\mu\phi\nabla^\mu\phi+4\nabla_\mu\nabla^\mu\phi-\frac{1}{12}\tilde{H}_{\mu\nu\rho}\tilde{H}^{\mu\nu\rho}\\
&+\frac{a}{8}R^{(-)}_{\mu\nu a}{}^bR^{(-)\mu\nu}{}_b{}^a+\frac{b}{8}R^{(+)}_{\mu\nu a}{}^bR^{(+)\mu\nu}{}_b{}^a\bigg) .  
\end{aligned}
\end{equation}
Here the modified field strength and the Chern-Simons three-form with torsion are defined by
\begin{equation}\label{tilde H}
\tilde{H}_{\mu\nu\rho}=H_{\mu\nu\rho}-\frac{3}{2}a\Omega^{(-)}_{\mu\nu\rho}+\frac{3}{2}b\Omega^{(+)}_{\mu\nu\rho}, \quad \Omega^{(\pm)}_{\mu\nu\rho}=\omega^{(\pm)}_{[\mu a}{}^b\partial_\nu\omega^{(\pm)}_{\rho]b}{}^a+\frac{2}{3}\omega^{(\pm)}_{[\mu a}{}^b\omega^{(\pm)}_{\nu b}{}^c\omega^{(\pm)}_{\rho]c}{}^a~\,,    
\end{equation}
while the modified spin connection and Riemann tensor are defined by 
\begin{equation}\label{omegapm}
\omega^{(\pm)}_{\mu a}{}^b=\omega_{\mu a}{}^b\pm\frac{1}{2}H_{\mu a}{}^b    
\end{equation}
and
\bea\label{RpmDef}
R^{(\pm)}_{\mu\nu a b} =R_{\mu\nu a b}\pm \mathcal{D}_{[\mu}H_{\nu]ab}-\frac{1}{2}H_{[\mu a }{}^cH_{\nu]cb}\,.   
\eea
The modified curvatures can also be expressed in terms of the modified connections as
\begin{equation}\label{R tensor Twist}
R^{(\pm)}_{\mu\nu ab}=\partial_\mu\omega^{(\pm)}_\nu{}^{ab}-
\partial_\nu\omega^{(\pm)}_{\mu}{}^{ab}+
\omega^{(\pm)}_{\mu}{}^{ad}\eta_{cd}\omega^{(\pm)}_\nu{}^{cb}    
-\omega^{(\pm)}_{\nu}{}^{ad}\eta_{cd}\omega^{(\pm)}_\mu{}^{cb}\,.
\end{equation}
Our conventions are summarized in the Appendix \ref{conventions}. 

The parameters $(a,b)$ in the action (\ref{rpm action}) take values $(-\alpha^\prime,-\alpha^\prime)$ for bosonic, $(-\alpha^\prime,0)$ for heterotic, and $(0,0)$ for the type II string. Note that, while for the type II string theories, the second line in (\ref{rpm action}) always vanishes, and the leading $\alpha^\prime$-corrections contribute at the order $\mathcal{O}({\alpha'}^3)$, the solutions of the heterotic and bosonic theories can be corrected already by the second line in (\ref{rpm action}). However, in the heterotic case, such corrections vanish for configurations with 
\bea\label{RminZero}
R^{(-)}_{\mu\nu\la\sigma}=0,
\eea
and such solutions will be analyzed in the next section.

\section{Geometries with vanishing twisted curvature}
\label{SecRmin}

Motivated by the previous discussion, in this section we will construct the most general solutions of the relations (\ref{RminZero}). First we note that in the absence of fluxes, the conditions (\ref{RminZero}) imply that the space is flat\footnote{Since (\ref{RminZero}) is a local condition, the space can still have a nontrivial topology, i.e., contain tori. We will focus on the local geometry and ignore potential periodic identifications.}. In this section we will show that the condition (\ref{RminZero}) gives equally strong constraints: it guarantees that the space is a product of group manifolds and flat space. After establishing this in section \ref{SecSubRmDerv}, we will analyze some physical properties of the resulting solutions in section \ref{SecSubRmEOM}.

\subsection{The most general geometries with $R^{(-)}=0$}
\label{SecSubRmDerv}

As the first step in analyzing the constraint (\ref{RminZero}),
we look at an expansion near some generic point. By choosing convenient coordinates and imposing a gauge on the two--form, we can expand the fields as
\bea\label{gBexpand}
g_{\mu\nu}=\eta_{\mu \nu}+\eps h_{\mu \nu}+\eps^2 h^{(2)}_{\mu\nu}+\dots,\quad
B_{\mu\nu}=\eps b_{\mu\nu}+\eps^2 b^{(2)}_{\mu\nu}+\dots
\eea
and treat $\eps$ as a small parameter. Substitution of these expansions into the expression (\ref{RpmDef}) for $R^{(-)}$ gives
\bea
R^{(-)}_{\mu\nu\rho\sigma}=\frac{\eps}{2}\left[\d_\nu\d_\rho h^{(-)}_{\mu\sigma}-\d_\mu\d_\rho h^{(-)}_{\nu\sigma}-\d_\nu\d_\sigma h^{(-)}_{\mu\rho}+
\d_\mu\d_\sigma h^{(-)}_{\nu \rho}\right]+O(\eps^2),\quad h^{(-)}_{\mu\nu}=
h_{\mu\nu}-b_{\mu\nu}\,.\nonumber
\eea
In particular, the metric and the Kalb--Ramond field can be decoupled by taking two linear combinations:		
\bea\label{RplusDecpl}
&&R^{(-)}_{\mu\nu\rho\sigma}-R^{(-)}_{\rho\sigma\mu\nu}=\eps\left[\d_\nu\d_\rho h_{\mu\sigma}-\d_\mu\d_\rho h_{km}-\d_\nu\d_\sigma h_{\mu\rho}+
\d_\mu\d_\sigma h_{\nu\rho}\right]+O(\eps^2)\,,\nn
&&R^{(-)}_{\mu\nu\rho\sigma}+R^{(-)}_{\rho\sigma\mu\nu}=\eps\left[\d_\nu\d_\rho b_{\mu\sigma}-\d_\nu\d_\sigma b_{\mu\rho}+\d_\mu\d_\sigma b_{\nu\rho}-\d_\mu\d_\rho b_{\nu \sigma}\right]+O(\eps^2)\,.
\eea
Equation (\ref{RminZero}) implies that both lines must vanish. The first line is the leading term in the expansion of the Riemann tensor around flat space, and it vanishes if and only if $h_{\mu\nu}$ is a pure gauge. Therefore by performing an $\eps$--dependent diffeomorphism, we can set $h_{\mu\nu}=0$ in (\ref{gBexpand}). Imposing equations (\ref{RminZero}) in the second line of (\ref{RplusDecpl}), we arrive at relations
\bea\label{Heqn1ordMain}
\d_\nu c_{\rho\mu\sigma}-\d_\mu c_{\nu\rho\sigma}=0,\quad \mbox{where}\quad c_{\rho\mu\sigma}=\d_\rho b_{\mu\sigma}+\d_\sigma b_{\rho\mu}+\d_\mu b_{\sigma\rho}\,.
\eea
Tensor $c_{\rho\mu\sigma}$ is the leading contribution to the $H$--field. The subsequent orders in perturbation theory are analyzed in the Appendix \ref{SecAppRmPert}, and the arguments presented there indicate that $c_{\rho\mu\sigma}$ are constant components subject to constraints
\bea\label{ZconstrEqn}
Z_{\mu\nu\rho\sigma}=c_{\mu\nu \alpha}c_{\rho\sigma\alpha}+c_{\mu\rho\alpha}c_{\sigma\nu\alpha}+c_{\mu\sigma\alpha}c_{\nu\rho\alpha}=0\,.
\eea
Note that the same constraint is satisfied by the structure constants of an arbitrary Lie group, so our result can be viewed as a local version of the Cartan--Schouten theorems \cite{Cartan1,Cartan2} which state that any simply connected manifold that admits a (non--symmetric) connection with zero--curvature is either a Lie group or
the 7-sphere\footnote{Since our torsion is exact, $H=dB$, and therefore closed, the case of $S^7$ is excluded from our analysis.}.

As further demonstrated in the Appendix \ref{SecAppRmPert}, in the physically most interesting case where $H$ has indices in a $d\le 6$ subspace, the constraints (\ref{ZconstrEqn}) imply that by performing rotations among $x$ coordinates\footnote{Recall that the expansions (\ref{gBexpand}) assume that $g_{\mu\nu}=\eta_{\mu\nu}$ in the leading order, and this leaves a freedom in rotating coordinates.}, one can split the directions into triplets with potentially nontrivial components
\bea\label{c123temp}
c_{123},\quad c_{456}. 
\eea
All other components of the field strength must vanish. In this basis, the full geometry (not only the field strength) splits into a product of three--dimensional subspaces which have the form
\bea\label{NSstart}
ds^2=f(r)(dx_1^2+dx_2^2+dx_3^2),\quad H=F(r) dx_1\wedge dx_2\wedge dx_3.
\eea
Here $r$ is the radial coordinate in the $(x_1,x_2,x_3)$ space. 

The perturbative analysis presented in the Appendix \ref{SecAppRmPert} demonstrates that the most general solution of the relation (\ref{RminZero}) in $d\le 6$ is the product of spaces (\ref{NSstart}), and it also gives the first few terms in the expansions of functions $(f,F)$. However, at this point,
we can just start from (\ref{NSstart}) and compute the exact expressions for the components of
$R^{(-)}_{\mu\nu\la\sigma}$. For example, 
\bea
R_{1212}^{(-)}&=&F^2 + \left[-2 + \frac{3 x_3^2}{r^2}\right]{\dot f}^{2} + 
 2 f \Big[\frac{\dot f}{r} + (1 - \frac{x_3^2}{r^2}) {\ddot f}\Big]=0,\nn
 R_{1231}^{(-)}&=&-\frac{3{\dot f}}{2r}\left[x_1F + \frac{x_2 x_3}{r} {\dot f}\right] + 
 \frac{1}{r}\left[x_1 {\dot F} + 
    \frac{x_2 x_3}{r} {\ddot f}\right]=0.
\eea
The last two equations form an overdetermined system for two functions of the radial coordinate, and the most general solution reads
\bea\label{RminSoln}
    f=\frac{c_2}{[1+(c_1 r)^2]^2}\,,\quad F=\frac{-4 c_2c_1}{(1+c_1^2 r)^3}\,,
\eea
where $c_1$ and $c_2$ are some integration constants. Eliminating the parameter $c_2$ by rescaling the coordinates $(x_1,x_2,x_3)$, we arrive at the final answer
\bea\label{NSfin}
ds^2=\frac{dx_1^2+dx_2^2+dx_3^2}{[1+(c\,r)^2]^2},\quad 
H=\frac{-4c}{[1+(c\,r)^2]^3} dx_1\wedge dx_2\wedge dx_3.
\eea
This geometry describes a three--dimensional sphere with the $H$ flux proportional to the volume form. Such spaces arise in string theory by taking near--horizon limits of NS5 branes \cite{CHS}, but it is important to note that we arrived at (\ref{NSfin}) by imposing the condition (\ref{RminZero}) rather than equations of motion. Although in this article we will focus on the solutions of (\ref{RminZero}) that can be written as products of the geometries (\ref{NSfin}) and flat spaces\footnote{In addition to products of two three--spheres, in six compact dimensions one also finds the geometry of $SO(4)\simeq SU(2)\otimes SU(2)/Z_2$, which has the metric (\ref{NSfinProd}) with additional global identifications.},
\bea\label{NSfinProd}
ds^2&=&\sum q_k^2 d\Omega_{3,(k)}^2+ds_{p,1}^2,\quad H=\sum q_k^3 d^3\Omega_{3,(k)}\,.
\eea
let us briefly comment on other solutions of equation (\ref{RminZero}). 

Since the relation (\ref{ZconstrEqn}) can be viewed as the Jacobi identity for a Lie algebra, the counterpart of the transition from (\ref{c123temp}) to (\ref{NSfinProd}) leads to the geometry of the corresponding Lie group, in agreement with the Cartan--Schouten theorem \cite{Cartan1,Cartan2}. As discussed in the Appendix \ref{SecAppRmPert}, the counterpart of the result (\ref{NSfinProd}) is
\bea\label{NSfinProdG}
ds^2&=&\sum_k q_k^2 \left[\sum_a e^a_{(k)}e^a_{(k)}\right]+ds_{p,1}^2,\quad 
H=\sum_k q_k^3 f^{(k)}_{abc}e^a_{(k)}\wedge e^b_{(k)}\wedge e^c_{(k)}\,,
\eea
where $e^a_{(k)}$ are frames for the manifold of simple Lie group $G_k$ with normalization
\bea\label{EframeMK}
de_{(k)}^a=f^{(k)}_{abc} e_{(k)}^b\wedge e_{(k)}^c
\eea
Equation (\ref{NSfinProd}) is the special case of (\ref{NSfinProdG}) for a product of $SU(2)$ factors. Note that while the shifted curvature appearing in (\ref{RminZero}) does not depend on the frames, spin connections do. The situation is analogous to flat space, which has a vanishing Riemann tensor, while the spin connections depend on the choice of frames, and in the most obvious option, $e^a_\mu=\delta^a_\mu$, they vanish as well. Since the spin connections $\omega^{(-)}$ and the corresponding Chern--Simons forms 
$\Omega^{(-)}$ prominently appear in the quantum corrections to the action (recall equations (\ref{rpm action}) and (\ref{tilde H})), it would be beneficial to select the frames where $\omega^{(-)}$ takes the simplest form. As outlined at the end of the Appendix \ref{SecAppRmPert}, the frames $e_{(k)}^b$ appearing in (\ref{NSfinProdG}), (\ref{EframeMK}) lead to vanishing twisted spin connections $\omega^{(-)}$, so they drastically simplify the structure of quantum corrections discussed later in this article. 

The results obtained in this subsection are qualitatively summarized in the second line of Table \ref{Table1}. The first line of that table represents the well known facts that vanishing Riemann curvature implies that the space is flat, which in turn implies that there exist frames with vanishing spin connections. The second line justified in this subsection shows very similar relations for $R^{(-)}$. This brings a natural question: are there special frames if three--spheres are replaced by $S^m$ or hyperbolic spaces? In the next section we will show that the natural counterpart of the condition $\omega^{(-)}=0$ is a requirement for the flat projections of the spin connections, ${\omega_c}^{ab}$, to be constant, and this uniquely fixes the frames up to rotation. This result is summarized in the last line of Table \ref{Table1}.

\begin{table}
\begin{center}
\begin{tabular}{|ccccc|} 
\hline
curvature&&geometry&&special frames\\
 \hline
 $R_{\mu\nu\la\sigma}=0$&$\Rightarrow$&$R^{m,n}$&$\Rightarrow$&$\omega=0$\\
 \hline
  $R^{(-)}_{\mu\nu\la\sigma}=0$&$\Rightarrow$&$R^{m,n}\times G_1\times \dots \times G_k$&$\Rightarrow$&$\omega^{(-)}=0$\\
  \hline
 &&$S^{m}$&$\Rightarrow$&${\omega_c}^{ab}=\,$const\\
 \hline
\end{tabular}
\end{center}
\caption{Summary of the relations between curvature, geometry, and special frames.}
\label{Table1}
\end{table}

To conclude this subsection, let us analyze the $\alpha'$ corrections to the geometries (\ref{NSfinProd}) and (\ref{NSfinProdG}). Interestingly, although these configurations were constructed by solving the condition (\ref{RminZero}), they satisfy the relation 
\bea\label{RplusAccid}
R^{(+)}_{\mu\nu\la\sigma}=0 
\eea
as well. This can be verified by a direct calculation or established using the following argument. Let us define two linear combinations generalizing the linearized expressions (\ref{RplusDecpl}):
\bea
\tilde{\mathcal{R}}^{(+)}_{\mu\nu\rho\sigma}\equiv R^{(+)}_{\mu\nu\rho\sigma}+
R^{(+)}_{\rho\sigma\mu\nu},\quad     
\tilde{\mathcal{R}}^{(-)}_{\mu\nu\rho\sigma}\equiv R^{(+)}_{\mu\nu\rho\sigma}-R^{(+)}_{\rho\sigma\mu\nu}\,.
\eea
Recalling the definitions (\ref{RpmDef}), we conclude that $\tilde{\mathcal{R}}^{(\pm)}$ can also be written in terms of $R^{(-)}$:
\bea
\tilde{\mathcal{R}}^{(+)}_{\mu\nu\rho\sigma}= R^{(-)}_{\mu\nu\rho\sigma}+
R^{(-)}_{\rho\sigma\mu\nu},\quad     
\tilde{\mathcal{R}}^{(-)}_{\mu\nu\rho\sigma}=- R^{(-)}_{\mu\nu\rho\sigma}+R^{(-)}_{\rho\sigma\mu\nu}\,.
\eea
Therefore, the geometries satisfying the condition (\ref{RminZero}) have vanishing  $\tilde{\mathcal{R}}^{(\pm)}$ and thus vanishing $R^{(+)}$. In other words, relation (\ref{RminZero}) implies (\ref{RplusAccid}). 

In the next subsection we will show that the geometries (\ref{NSfinProd}),  (\ref{NSfinProdG}) satisfy the equations of motion coming from the action  (\ref{rpm action}), and that with appropriate field redefinitions, the supergravity solution is exact. This property of the NS5 brane geometries in heterotic theories has been known for a long time \cite{CHS}, but the special role played by the constraint (\ref{RminZero}) is a new feature of our analysis.

\subsection{Equations of motion with $\alpha'$ corrections}
\label{SecSubRmEOM}

In the previous subsection we found the most general solution of the relations (\ref{RminZero}), which guarantee that the second line of the action (\ref{rpm action}) does not contribute to equations of motion in the heterotic theory, where $b=0$. However, this does not fully eliminate the possibility of nontrivial $\alpha'$ corrections due to the shift in the three--form field strength. In this subsection we will show that the spaces (\ref{NSfinProd}) do satisfy the equations of motion coming from the truncated version of (\ref{rpm action}),
\bea\label{RpmTrunc}
S&=&\int dx\sqrt{-g}e^{-2\phi}\bigg[R-4\nabla_\mu\phi\nabla^\mu\phi+4\nabla_\mu\nabla^\mu\phi-\frac{1}{12}\tilde{H}_{\mu\nu\rho}\tilde{H}^{\mu\nu\rho}\bigg],\\   
&&\tilde{H}_{\mu\nu\rho}=H_{\mu\nu\rho}-\frac{3}{2}a\Omega^{(-)}_{\mu\nu\rho}\,,\nonumber
\eea
once an appropriate dilaton is turned on. 

We begin with observing that the action (\ref{RpmTrunc}) differs from its supergravity approximation (which corresponds to $a=0$) only through modification of the Bianchi identity:
\begin{equation}\label{Bianchi}
d\tilde{H}=-\alpha^\prime\operatorname{Tr}[R\wedge R],\quad~~~~\text{where} ~~~R^{ab}=\frac{1}{2!}R^{ab}{}_{\mu\nu}dx^\mu\wedge dx^\nu   .  
\end{equation}
Trace in the expression above ensures that different spheres from (\ref{NSfinProd}) decouple in these identities, 
then expressions $\operatorname{Tr}[R\wedge R]$ produce four--forms on three dimensional spaces, which vanish kinematically. Therefore, one can define a shifted field ${\tilde B}$ by the relation
\bea\label{Bshift}
(d{\tilde B})_{\mu\nu\rho}=H_{\mu\nu\rho}-\frac{3}{2}a\Omega^{(-)}_{\mu\nu\rho}\,.
\eea
Then the geometry 
\bea\label{NsProdSugra}
ds^2=\sum q_k^2 d\Omega_{3,(k)}^2+ds_{p,1}^2,\quad d{\tilde B}=\sum q_k^3 d^3\Omega_{3,(k)}
\eea
satisfies the {\it supergravity} equations of motion coming from the action (\ref{RpmTrunc}), provided that the dilaton is a linear function of the transverse coordinates contained in $ds_{p,1}^2$, e.g., 
\bea
\phi=\left[\sum 2c_k\right]x\,,
\eea
where $x$ is one of the transverse directions. 
This is a well--known CHS solution describing the near--horizon limit of NS5 banes \cite{CHS}. 

Note that while the metric and the shifted field ${\tilde B}$ are exact, the original field $B$ may or may not receive corrections, depending on the prescription for doing quantum computations. In the DFT formalism discussed here, the prescription is encoded in frames which lead to $\Omega^{(-)}_{\mu\nu\rho}$. For example, choosing the widely used frames 
corresponding to the metric (\ref{NSfin}) of the 3--sphere,
\bea\label{FrameTriv}
e^a=\frac{dx^a}{1+(c\, r)^2}\,,
\eea
we arrive at the expression for the original field strength:
\bea \label{redefined H order 1}
(dB)_{123}=H_{123}=\frac{4c}{[1+(c\, r)^2]^3}-\frac{4c^3}{[1+(c\, r)^2]^2}\alpha^\prime.
\eea
Other choices of frames lead to different relations between $B$ and ${\tilde B}$, and in the next subsection we will introduce the unique special frames which ensure that $B={\tilde B}$, so that the Kalb--Ramond field does not receive quantum corrections.  

Interestingly, the redefinition (\ref{redefined H order 1}) can be extended to higher orders in $\alpha'$. While the full effective action of string theory is not known, the extension of (\ref{rpm action}) to the next order was recently obtained in \cite{2206.10640}, and for the heterotic string it reads
\bea
S&=&\int dx\sqrt{-g}e^{-2\phi}\left[R-4\nabla_\mu\phi\nabla^\mu\phi+4\nabla_\mu\nabla^\mu\phi-\frac{1}{12}\tilde{H}_{\mu\nu\rho}\tilde{H}^{\mu\nu\rho}
+\frac{a}{8}\hat{R}^{(-)}_{\mu\nu a}{}^b\hat{R}^{(-)\mu\nu}{}_b{}^a\right.\nonumber\\
&&\left.+\mathcal{O}(\alpha^{\prime 3})\right]\,.
\eea
While this action looks identical to  (\ref{rpm action}), the modified field strength is more complicated:
\begin{equation}\label{hat H}
\begin{aligned}
\hat{H}_{\mu\nu\rho}&=H_{\mu\nu\rho}-\frac{3a}{2}\Omega^{(-)}_{\mu\nu\rho}+\frac{9a^2}{8}\bigg(\partial_{[\mu}(\Omega^{(-)}_\nu{}^{\sigma\beta}\omega_{\rho]\sigma\beta}^{(-)})+\Omega^{(-)}_{[\mu}{}^{\sigma\beta}R_{\nu\rho]\sigma\beta}^{(-)}\bigg)+\mathcal{O}(\alpha^{\prime 3}).
\end{aligned}    
\end{equation}
Once the condition (\ref{RminZero}) is imposed, the geometry takes the product form (\ref{NSfinProd}). Then modified Bianchi identity for the field $\hat{H}$ defined by (\ref{hat H}) is also satisfied kinematically, so once again, the quantum corrections can be absorbed by a field redefinition. Thus, at least up to 
$\mathcal{O}(\alpha^{\prime 3})$ corrections, the geometry (\ref{NsProdSugra}) written in terms of $\hat{H}$ is exact, and the original $B$ field can be recovered from 
\begin{equation}
\begin{aligned}
dB_{123}=H_{123}=\frac{4c}{[1+(c\,r)^2]^3}-\frac{4c_1^3}{[1+c_1^2r^2]^2}\alpha^\prime+ 
\frac{4c^5(3+(c\, r)^2)}{[1+(c\, r)^2]^2} \alpha^{\prime2}\,.
\end{aligned}    
\end{equation}
The explicit expression for the $B$ field is not very illuminating, and we present it just for completeness
\bea\label{BfieldAphShift}
B=\left[\frac{r[(c\,r)^2-1]}{2c[1+(c\, r)^2]^2}+\frac{2c\,r\,\alpha'}{1+(c\, r)^2}+
\frac{4c^5\,r^3\,(\alpha')^2}{1+(c\, r)^2}+\left(\frac{1}{2c^2}-2\alpha'\right)\arctan(c\,r)\right]d\Omega_2\,.
\eea
We used the spherical coordinates in the three--dimensional space spanned by $(x_1,x_2,x_3)$.

So far we have analyzed the $\alpha'$ corrections in (\ref{hat H}) in the frames (\ref{FrameTriv}) which are traditionally used for the three--sphere in the string theory literature. However, as we discussed in the previous subsection, the frames defined by 
\bea
{\tilde e}^a=\frac{1}{c}\mbox{tr}(g^{-1}dg T^a)
\eea
are more natural from the group theoretic perspective, and they lead to vanishing 
$\omega^{(-)}$ and $\Omega^{(-)}$, i.e., to $\hat{H}_{\mu\nu\rho}=H_{\mu\nu\rho}$ and to replacement of (\ref{BfieldAphShift}) by the undeformed expression
\bea\label{BfieldAphShiftGrp}
B=\frac{r[(c\,r)^2-1]}{2c[1+(c\, r)^2]^2}\, d\Omega_2\,.
\eea
For completeness, we present the explicit expressions for the frames ${\tilde e}^a$ in terms of coordinates used in (\ref{FrameTriv}):
\begin{equation}\label{new frames}
{\tilde e}_{m}{}^a=f(r)\begin{bmatrix}
{1-c^2(r^2-2x_1^2)}&{2c(c\,x_1x_2+x_3)}&{2c(c\,x_1x_3-x_2)}\\
{2c(c\,x_1x_2-x_3)}&{1-c^2(r^2-2x_2^2)}&{2c(c\,x_3x_2+x_1)}\\
{2c(c\,x_1x_3+x_2)}&{2c(c\,x_3x_2-x_1)}&{1-c^2(r^2-2x_3^2)}
\end{bmatrix}\,,
\end{equation}
where
\bea
{\tilde e}^a={\tilde e}_{n}{}^a dx^m,\quad f(r)=\frac{1}{[1+(c\,r)^2]^2},\quad r^2=x_m x_m\,.
\eea
This concludes our discussion of the $\alpha'$ corrections in the heterotic string.

Let us now make some comments about the bosonic string, where the parameters $(a,b)$ in the action 
(\ref{rpm action}) take values $(-\alpha^\prime,-\alpha^\prime)$. While we have been imposing the condition (\ref{RminZero}), as demonstrated in the previous subsection, the most general solution of this relation, (\ref{NSfinProd}), also satisfies the relation (\ref{RplusAccid}). Therefore, the second line in (\ref{rpm action}) vanishes for such geometries regardless of values of $a$ and $b$, so we get a supergravity action with ${\tilde H}$ that satisfies the standard Bianchi identities (see the discussion around equation (\ref{Bianchi})). We conclude that the geometry (\ref{NsProdSugra}) solves equations of motion for the bosonic strings as well, but in this case, the counterpart of the re-definition (\ref{redefined H order 1}) is
\bea\label{redefined H order 1 bosonic}
H_{123}=\frac{4c}{[1+(c\, r)^2]^3}-\frac{8c^3}{[1+(c\, r)^2]^2}\alpha^\prime.
\eea
This three--form  $H$ is exact, and it can be integrated to produce a counterpart of (\ref{BfieldAphShift}).
To arrive at the relation (\ref{redefined H order 1 bosonic}), we again used the obvious frames (\ref{FrameTriv}), but as clear from the expressions (\ref{tilde H}) and (\ref{hat H}), the details of the shifts depend on the choice of frames. While in the heterotic case, the frames (\ref{new frames}) ensure disappearance of $\alpha'$ corrections for the geometry (\ref{NSfinProd}), in the bosonic case, the shift in the $H$ field (\ref{tilde H}) still receives a nontrivial contribution from $\Omega^{(+)}_{\mu\nu\rho}$. Then the exact solution (\ref{NsProdSugra}) for the shifted field leads to
\bea
H_{123}=\frac{4c}{[1+(c\,r)^2]^3}-\frac{32c^3}{[1+(c\,r)^2]^3}\alpha'.
\eea
The resulting three--form $H$ is exact, and the corresponding Kalb--Ramond potential has the form $B=h(r)d\Omega_2$.

Our analysis of quantum corrections is based on the condition (\ref{RminZero}), which naturally arises in the DFT prescription. Unfortunately, recently some difficulties have been encountered in extending this prescription beyond to the $\mathcal{O}(\alpha^{\prime 3})$ order \cite{AlpObstr}, and it would be nice to see whether these problems can be overcome at least for the backgrounds (\ref{NSfinProd}). We leave exploration of this direction for future work.

\section{Special frames for spheres and hyperboloids}
\label{SecFrameSn}

As discussed in the previous section, the condition (\ref{RminZero}) implies that the space--time is a product of group manifolds and flat space:
\bea\label{SpaceSk}
{\cal M}=R^{p,1}\times \prod_k G_k\,.
\eea
Furthermore, it is possible to choose frames which give simple spin--connections:
\bea
{\omega_\mu}^{ab}={H_\mu}^{ab}\,.
\eea
In particular, for spaces (\ref{SpaceSk}) this implies that the projections of spin--connections to frames are constant:
\bea\label{OmegaConst}
{\omega_c}^{ab}\equiv e_c^\mu {\omega_\mu}^{ab}=\mbox{const}\,.
\eea
In this section we will demonstrate that the conditions (\ref{OmegaConst}) can be satisfied for any product of spheres, hyperboloids, AdS, and flat spaces. For a flat space, one can always choose constant frames, which lead to ${\omega_c}^{ab}=0$. Let us now consider $p$--dimensional spheres and hyperboloids.

A metric on a $d$--dimensional space of constant curvature can be written as
\bea\label{SphereWpar}
ds^2=\frac{1}{(1+w^2 r^2)^2}dy_s dy_s,\quad r^2=y_s y_s\,.
\eea
The Riemann tensor satisfies the relation
\bea
R_{\mu\nu\alpha\beta}=4w^2\left[g_{\mu\alpha}g_{\nu\beta}-g_{\nu\alpha}g_{\mu\beta}\right],
\eea
where $w$ is real for spheres and imaginary for hyperboloids. For flat space ($w=0$), the natural frames are $e^a_s=\delta^a_s$, and for the geometry (\ref{SphereWpar}), one can look at an expansion in powers of $w$:
\bea\label{frameWexpnd}
e^a=dy^a+w A_{abc}\,y^b dy^c+w^2 B_{abcd}\,y^b y^c dy^d+\dots
\eea
Let us summarize the properties of this expansion.
\begin{enumerate}[(a)]
\item The requirement that (\ref{frameWexpnd}) are frames leads to constraints $A_{abc}=-A_{cba}$ and determines some coefficients $B_{abcd}$ in terms of $A_{abc}$.
\item The requirement (\ref{OmegaConst}) imposed up to the second order in $w$ determines all remaining $B_{abcd}$ and leads to a system of quadratic relations for parameters $A_{abc}$.   For example, in four dimensions, there are $4\times 24=96$ independent coefficients in front of various $y_s$ in ${\omega_c}^{ab}$, and they lead to $36$ quadratic relations. 
\item Although for $d\ge 4$ it is impossible to solve the quadratic relations from part (b) while preserving the $SO(d)$ symmetry, there exists a unique solution that preserves the rotations between $(y_2,\dots y_d)$ coordinates\footnote{Strictly speaking, there is a freedom in choosing one sign, so in addition to (\ref{SoldOmegPert}), there is one more solution obtained by flipping the sign of $w$.}:
\bea\label{SoldOmegPert}
e^1&=&dy_1-2iw\rho d\rho+w^2\left[2y_1 \rho d\rho+(\rho^2-(y_1)^2)dy_1\right]+O(w^3)\\
e^a&=&\left[1-w^2 r^2\right]dy_a+2iwy_a(1+iwy_1)dy_1+2w^2y_a \rho d\rho+O(w^3),\quad a>1.\nonumber
\eea
Here $\rho=\sqrt{(y_2)^2+\dots+(y_d)^2}=\sqrt{r^2-(y_1)^2}$. In particular, this solution has only $2(d-1)$ nontrivial coefficients $A_{abc}$:
\bea\label{SoldOmegPertAA}
A_{122}=\dots=A_{1dd}=-2i,\quad A_{212}=\dots=A_{d1d}=2i\,.
\eea
\item Perturbing the coefficients $A_{abc}$ from (\ref{frameWexpnd}) around solution (\ref{SoldOmegPertAA}) (i.e., relaxing the requirement of the $SO(d-1)$ invariance under rotations in the $(y_2,\dots y_d)$ subspace) and substituting the result into the quadratic equations from part (b), one finds a system of linear equations for the perturbations. For example, in $d=4$, this system has six linearly--independent solutions, which can be interpreted as infinitesimal $SO(4)$ rotations of the coordinates and frames:
\bea\label{SoldOmegPertRot}
{e'}^a={R^a}_b e^b,\quad y'_a={R^a}_b y_b\,.
\eea
A similar interpretation of the perturbations works in all dimensions. Therefore, we conclude that the perturbative expansion (\ref{SoldOmegPert}) is unique up to rotations (\ref{SoldOmegPertRot}). 
\end{enumerate}
To summarize, we have demonstrated that in $d\ge 4$ dimensions, any perturbative expansion of the frames (\ref{frameWexpnd}) that satisfies the condition (\ref{OmegaConst}) can be transformed into the form (\ref{SoldOmegPert}) by an $SO(d)$ rotation (\ref{SoldOmegPertRot}). Expression (\ref{SoldOmegPert}) gives the unique solution of the conditions (\ref{OmegaConst}) with $SO(d-1)$ symmetry, and exact frames with such symmetry must have the form
\bea
e^1&=&f_1 dy_1+f_2 r dr,\nn
e^a&=&f_3 dy_a+y_a[f_4 dy_1+f_5 r dr],\quad a>1,
\eea 
where $f_k$ are functions of $y_1$ and $r$. Evaluating the spin--connection and imposing the constraint (\ref{OmegaConst}), we find the {\it unique} solution for the frames\footnote{As mentioned earlier, strictly speaking, there are two solutions: (\ref{AnswFramesWconst}) and its counterpart with a flipped sign of $w$.}
\bea\label{AnswFramesWconst}
e^1&=&f\Big[[1+(w r)^2]dy_1-2iwr(1-iwy_1) dr\Big],\nn
e^a&=&\frac{dy_a}{1+(wr)^2}+2iwfy_a[dy_1-iw r dr],\quad a>1,\\
f&=&\frac{1}{[1+(w r)^2][1-2iwy_1-(wr)^2]}\,.\nonumber
\eea
The non--zero components of the frames ${\omega_c}^{ab}$ are
\bea\label{AnswOmegWconst}
{\omega_2}^{12}=\dots={\omega_d}^{1d}=2iw,\quad
{\omega_2}^{21}=\dots={\omega_d}^{d1}=-2iw\,.
\eea
The discussion above focused on $d\ge 4$, and in three dimensions, the frames 
(\ref{new frames}) give a more symmetric result: ${\omega_a}_{bc}=4c\eps_{abc}$. In two dimensions, one can select holomorphic frames:
\bea\label{S2holom}
e^+=\frac{1+iw{\bar z}}{1+iw{z}}\frac{dz}{1+w^2 z{\bar z}},\quad 
e^-=\frac{1+iw{z}}{1+iw{\bar z}}\frac{d{\bar z}}{1+w^2 z{\bar z}},\quad
{\omega_+}^{+-}={\omega_-}^{-+}=-iw.
\eea
Derivation of this result is presented in the Appendix \ref{AppFrame2d}. 

To summarize, in this section we showed that every sphere and hyperboloid admits a set of frames that lead to constant spin connections (\ref{OmegaConst}), i.e., we justified the last line in Table \ref{Table1}. It is natural to ask an inverse question: does the condition (\ref{OmegaConst}) imply that the geometry is a product of spheres, hyperboloids and flat space? While we will not attempt to answer this question in full generality, in the Appendix \ref{AppFrame2d} we demonstrate that in two dimensions, the only spaces admitting constant frames (\ref{OmegaConst}) are $(S^2,H_2,R^2)$ and their analytic continuations, such as $R^{1,1}$. It would be interesting to extend this result to higher dimensions.

\section{Geometries with NS--NS fluxes}
\label{SecNSsoln}

Let us go back to the action (\ref{rpm action}) and analyze the $\alpha'$ corrections to some well--known solutions. Specifically, since the action (\ref{rpm action}) contains only the NS--NS fluxes, it is natural to focus on the NS5 branes and fundamental strings. We will show that the geometries produced by vibrating fundamental string don't receive $\alpha'$ corrections in the DFT prescription due to some nontrivial cancellations, while the geometries of the NS5 branes are corrected away from the near--horizon regime. We will consider these solutions in two separate subsections. 

\subsection{NS5 branes}

As demonstrated in section \ref{SecRmin}, the near horizon geometry of NS5 branes does not receive any quantum corrections from the action (\ref{rpm action}) in either heterotic or bosonic string theory, apart from the frame--dependent field redefinitions. It is natural to ask whether this property persists for the asymptotically--flat configurations of the NS5 branes\footnote{Note that the arguments of section \ref{SecRmin} imply that corrections to the supergravity solution describing NS5 branes beyond the near--horizon limit produce non-vanishing contributions to individual terms in the action (\ref{rpm action}). In this section, we are focusing on exploring potential cancellations between individual terms for NS5 branes and for fundamental strings.}. In this case, the supergravity solution is given by \cite{CHS}
\bea\label{CHSgeom}
ds^2&=&\eta_{ij}dx^i dx^j+H(r)\delta_{mn}dx^mdx^n,\\
\hat{H}^{(3)}_{mnp}&=&-\frac{1}{2}\epsilon_{mnpq}\partial_q H(r),\quad \phi=\frac{1}{2}\log H(r),\\
H(r)&=&h_0+\frac{Q}{r^2},\quad r^2=x_mx^m.
\eea
The charge $Q=n \alpha^\prime$ is associated with the modified $H$-flux,
\begin{equation}
Q=\frac{-1}{2\pi \alpha^\prime}\int_{S^3}\hat{H}^{(3)}\,.    
\end{equation}
The solution (\ref{CHSgeom}) is known to be exact at least up to the order $\mathcal{O}(\alpha^\prime)$, assuming that $n$ is kept fixed when $\alpha'$ is sent to zero \cite{CHS}. From the supergravity perspective, it might also be interesting to look at corrections to solutions with fixed $Q$. A direct evaluation of contributions to the curvature gives
\bea
R_{\mu\nu\rho\sigma}^{(+)}=\frac{2Qh_0}{r^2(Q+h_0r^2)^2} A_{\mu\nu\rho\sigma}\,,
\eea
where
\bea
A_{\mu\nu\rho\sigma}=2\bigg[\epsilon_{\mu\nu\lambda\delta}\epsilon_{\rho\sigma}{}^{\lambda\gamma}(x^\delta x_\gamma-\frac{1}{4}\delta^\delta_\gamma
x^\alpha x_\alpha)-\frac{x^\lambda}{32}\big(\epsilon_{\mu\nu\rho\lambda} x_\sigma-\epsilon_{\mu\nu\sigma\lambda} x_\rho-\epsilon_{\mu\lambda\rho\sigma} x_\nu+\epsilon_{\nu\lambda\rho\sigma} x_\mu\big)\bigg].\nonumber  
\eea 
This implies that even in the mini-superspace approach, where instead of varying with respect to all fields in (\ref{rpm action}), one takes variations only with respect to function $H$ appearing in the ansatz (\ref{CHSgeom}), this geometry receives nontrivial corrections, unless the constant $h_0$ vanishes, i.e., unless one takes the near--horizon limit of the five--branes.

\subsection{Fundamental strings}
\label{SecSubF1}

Let us now consider another important example of geometries with two--form fluxes, the solutions describing fundamental strings. The sigma models corresponding to such systems are known as the chiral null models \cite{CNMw1,CNMw2}, and their worldsheet action is given by\footnote{One can also consider more general configurations where function $K$ depends on the $u$ coordinate as well \cite{CNMw1,CNMw2}, and the analysis presented in this section applies to them as well.}
\bea\label{CNMstart}
S=\frac{1}{\pi\alpha^\prime}\int d^2z\big[F(x)\partial u\Bar{\partial} v+F(x)K(x)\partial u\Bar{\partial} u+\partial x_i\Bar{\partial}x^i  +\alpha^\prime\mathcal{R}\phi(X)\big]\,.    
\eea
As demonstrated in \cite{CNMw1,CNMw2}, in a certain prescription for computing quantum corrections, the solution (\ref{CNMstart}) is exact, as long as functions $F^{-1}$ and $K$ are harmonic and the dilaton is related to $F$:
\bea\label{CNMstart1}
\nabla^2 \frac{1}{F}=0,\quad \nabla^2 K=0,\quad \phi=\phi_0+\frac{1}{2}\ln F\,.
\eea
Since quantum corrections are prescription--dependent, it is interesting to see whether the system (\ref{CNMstart})--(\ref{CNMstart1}) satisfies the equations of motion coming from the DFT--inspired action (\ref{rpm action}). 

Rather than deriving all equations of motion from the Lagrangian $L=L^{(0)}+L^{(1)}$ given by (\ref{L0})--(\ref{L1}), we employ a minisuperspace approach of varying with respect to several relevant functions. Specifically, we consider an extended version of the chiral null model,
\bea\label{CNMmetr}
ds^2&=&F_1[du dv+K du^2]+dx_idx_i,\nn
B&=&F_2\,du\wedge dv,\quad \phi=\phi_0+\frac{1}{2}\ln F_3,
\eea
evaluate the Lagrangians  (\ref{L0})--(\ref{L1}) in terms functions $(F_1,F_2,F_3,K)$, and take variations with respect to them. In particular, the configurations which contain only $K$ (plane waves) don't contribute in the minisuperspace approach since they produce $R_{uuij}$ which contracts to zero. For $K=0$, straightforward but tedious calculations lead to the final results 
\bea\label{LagrCNMms}
L^{(0)}&=&\bigg[-\frac{2}{F_3}(\partial^i\partial_iF_1)+
\frac{1}{2F_1F_3}[(\nabla F_1)^2+(\nabla F_2)^2]-\frac{3F_1}{F_3^3}(\nabla F_3)^2\nn
&&+\frac{2F_1}{F_3^2}(\partial_i\partial^iF_3)+\frac{2}{F_3^2}(\partial_iF_1)(\partial^iF_3)\bigg],\nn
L^{(1)}&=&\frac{a+b}{4F_1^3F_3}\bigg[[(\partial_iF_1)(\partial^iF_2)]^2-
\frac{7}{2}(\nabla F_1)^2(\nabla F_2)^2+\frac{3}{4}(\nabla F_1)^4
+\frac{7}{4}(\nabla F_2)^4\\
&&-{2F_1}(\partial^i\partial^jF_1)\big[2(\partial_iF_2)(\partial_jF_2)+(\partial_iF_1)(\partial_jF_1)-F_1(\partial_i\partial_jF_1)\big]\nn
&&-{2}{F_1}(\partial_i\partial_jF_2)\big[F_1(\partial^i\partial^jF_2)-3(\partial^iF_2)(\partial^jF_1)\big]\nonumber
\bigg]\,.
\eea
Interestingly, these expressions vanish once the relation $F_1=F_2=F_3=F$ and equations (\ref{CNMstart1}) are imposed. Variations of (\ref{LagrCNMms}) with respect to $(F_1,F_2,F_3)$ 
are calculated in the Appendix \ref{SecAppCNM}, and the contributions from $L^{(1)}$ vanish once the relation $F_1=F_2=F_3=F$ and equations (\ref{CNMstart1}) are imposed. Therefore, the chiral null model is exact even in the DFT prescription, which is different from the original one used in \cite{CNMw1,CNMw2}. 
 
\section{Gravitational waves with $\alpha'$ corrections}
\label{SecGW}

In sections \ref{SecRmin} and \ref{SecNSsoln} we analyzed $\alpha^\prime$-corrections to several exact solutions of supergravity. Only a limited number of such solutions are known, and generically even in supergravity one must rely either on numerical analysis or on perturbative techniques. Perturbative solutions of the Einstein's equation, generally known as gravitational waves, play important roles in understanding of black hole physics, AdS/CFT correspondence, analysis of stability of exact solutions, etc., so it is interesting to study them in the presence of $\alpha'$ corrections as well. In this section we will do so assuming that the unperturbed background geometry is flat.

While combining the weak gravitational waves with $\alpha'$ corrections, one faces an interesting problem with the order of limits. Both perturbative gravitational waves and $\alpha'$ corrections are expected to be small, so one can write the expansion of the metric as
\bea
g_{\mu\nu}=\eta_{\mu\nu}+\eps h^{(1)}_{\mu\nu}+\eps^2 h^{(2)}_{\mu\nu}+\alpha'\eps h^{(1,1)}_{\mu\nu}+\dots
\eea
Here $\eps$ is a formal parameter controlling the strength of the wave. While the second term in the right hand side of the last expression always gives the leading contribution to the perturbation, the last two terms may compete with each other, depending on the relation between $\alpha'$ and $\eps$. In this section we will consider the limit where 
$\eps$ is taken to zero first, i.e., we will study linearized equations for gravitational waves governed by the action (\ref{rpm action}). 

In the absence of matter fields, the Lagrangian (\ref{L0})--(\ref{L1}) becomes
\begin{equation}\label{L0new}
L=R-\frac{(a+b)}{8}R_{\mu\nu\rho\sigma}R^{\mu\nu\rho\sigma}\,,
\end{equation}
and the corresponding equations of motion at the linear order are 
\bea\label{EomGWspec}
R_{\mu\nu}-\frac{1}{2}g_{\mu\nu}R-\frac{a+b}{4}\big( R_{,\mu\nu}-2 R_{\mu\nu}{}_{,\alpha}{}^\alpha\big)=0\,.  
\eea
It is instructive to consider a more general effective action
\bea\label{EomGWgenAct}
S=\int \sqrt{-g}dx\bigg[R+a_1R_{\mu\nu\rho\sigma}R^{\mu\nu\rho\sigma}+a_2R_{\mu\nu}R^{\mu\nu}+a_3R^2\bigg]\,,    
\eea
with free coefficients $(a_1,a_2,a_3)$. The linearized equations of motion generalize the expressions (\ref{EomGWspec}):
\bea\label{EomGWgen}
&&R_{\mu\nu}-\frac{1}{2}g_{\mu\nu}R\\
&&\quad=-2a_1\big[R_{,\mu\nu}-2R_{\mu\nu}{}_{,\alpha}{}^\alpha\big]-
a_2\big[R_,{}_{\mu\nu}-R_{\mu\nu}{}_{,\lambda}{}^\lambda-
\eta_{\mu\nu}R_{\alpha\beta}{}^{,\alpha\beta}\big]-2a_3\big[R_{,}{}_{\mu\nu}-
R_{,\alpha}{}^\alpha\eta_{\mu\nu}\big]\,.\nonumber
\eea
A comprehensive study of these equations, as well as their extensions to more general theories involving cubic and quartic terms in curvature, is performed in \cite{Viser16}. In contrast to that approach which contained sources, we focus on homogeneus equations and solve them algebraically in terms of plane wave. 

Substituting the expansion 
\begin{equation}\label{flat space perturbed}
g_{\mu\nu}=\eta_{\mu\nu}+\eps h_{\mu\nu}
\end{equation}
into the general expression for the Riemann tensor and expanding the result to the first order in $\eps$, we arrive at the well--known relation
\bea
 R_{\mu\nu\rho\sigma}=\frac{\eps}{2}(h_{\mu\sigma,\nu\rho}+h_{\nu\rho,\mu\sigma}-h_{\nu\sigma,\mu\rho}-h_{\mu\rho,\nu\sigma}),   
\eea
which has been extensively used for studying gravitational waves in flat space. In contrast to this standard analysis, which requires the left hand side of (\ref{EomGWgen}) to vanish, we will now take into account the quantum corrections coming from the right hand side and explore their effect on the dispersion relations. Specifically, starting from an ansatz for plane waves,
\bea
h_{\mu\nu}=\Tilde{h}_{\mu\nu}e^{ikx}\,,  
\eea
and imposing the gauge condition $\d^\mu h_{\mu\nu}=0$, we arrive at a set of algebraic equations
\bea\label{GWdispers}
k^2\left[\frac{1}{2}\big[\tilde{h}_{\mu\nu}-\tilde{h}\Pi_{\mu\nu}\big]+k^2\bigg\{\bigg[2a_1+\frac{a_2}{2}\bigg]\tilde{h}_{\mu\nu}+
\tilde{h}\bigg[\frac{a_2}{2}+2a_3\bigg]\Pi_{\mu\nu}\bigg\}\right]=0.
\eea
Here we have defined a projector $\Pi_{\mu\nu}$:
\bea
\Pi_{\mu\nu}=\eta_{\mu\nu}-\frac{k_\mu k_\nu}{k^2}:\qquad \Pi_{\mu\nu}\Pi^{\nu\sigma}=\Pi_\mu^\sigma,\quad
k^\mu \Pi_{\mu\nu}=0.
\eea
To write the relation (\ref{GWdispers}) in a compact form using this projector, we implicitly assumed that $k^2$ doesn't vanish, and in the degenerate case, equation (\ref{GWdispers}) is replaced by its limit,
\bea\label{GWdispersK0}
k^2=0:\quad k_\mu k_\nu {\tilde h}=0.
\eea
To find the dispersion relation for the wave vector $k$ in the generic case (\ref{GWdispers}), we take the trace of equation (\ref{GWdispers})
\bea\label{GWdispersK}
k^2\big[1-\xi k^2\big]\tilde{h}=0,\quad \xi=\frac{[4a_1+a_2+(d-1)(a_2+4a_3)]}{d-2} \,.
\eea
This relation has three types of solutions
\begin{enumerate}[(a)]
\item If $k^2=0$, then equation (\ref{GWdispersK0}) implies that ${\tilde h}=0$, and there are no additional constraints. These are the standard gravitational waves familiar from general relativity, and as expected, the excitations are traceless. Note that such waves do not receive quantum corrections, at least at the order 
$\mathcal{O}(\alpha')$.
\item If $\tilde{h}=0$, then equation (\ref{GWdispers}) becomes
\bea
k^2(\tilde{h}_{\mu\nu})\bigg[\bigg(2a_1+\frac{a_2}{2}\bigg)k^2 +\frac{1}{2}\bigg] =0,  
\eea
Assuming that $k^2$ does not vanish (this case was covered in option (a)), we arrive at the dispersion relation
\bea\label{GWnstDsp1}
k^2=-\frac{1}{4a_1+{a_2}}\,.
\eea
This branch appears due to quantum corrections, and $k^2$ goes to infinity in the classical limit. 
\item The final option is 
\bea\label{GWnstDsp2}
k^2=\frac{1}{\xi}\,,
\eea
This expression is positive in both bosonic and heterotic cases, and it goes to infinity in the classical limit. For this dispersion relation, equation (\ref{GWdispers}) gives
\bea
\tilde{h}_{\mu\nu}=\frac{1-(4a_3+a_2)k^2}{1+(4a_1+a_2)k^2}\,\tilde{h}\,\Pi_{\mu\nu}=
\frac{1}{d-1}\,\tilde{h}\,\Pi_{\mu\nu}\,.
\eea
Here we used the expression (\ref{GWdispersK}) for $\xi$. As a consistency check, we observe that the trace of the last equation gives an identity ${\tilde h}={\tilde h}$.
\end{enumerate}
To summarize, in this section we have analyzed the equations for gravitational waves governed by a generic action quadratic in curvature (\ref{EomGWgenAct}). We found that the standard plane waves don't receive quantum corrections, but two new branches appear, and they have non--standard dispersion relations (\ref{GWnstDsp1}) and (\ref{GWnstDsp2}). In both cases, the mass shell is pushed to infinity in the classical limit. It would be interesting to extend this analysis to quantum corrections at higher orders in $\alpha'$, but such discussion is outside the scope of this article which focuses on exploration of the truncated action (\ref{rpm action}).

\section{T duality and $\alpha'$ corrections}
\label{SecTdual}

So far we have focused on constructing solutions of equations of motion coming from the action (\ref{rpm action}) by using direct calculations. Once some such solutions are known, one can construct new ones by performing purely algebraic manipulations, such as dualities and shifts. Since the action (\ref{rpm action}) is invariant under T dualities, such operations are guaranteed to map solutions into solutions. We will focus on the situations when the original configuration does not receive quantum corrections. Then the solution produced by an abelian T duality will be exact as well, while application of a non--abelian duality generates nontrivial dependence on $\alpha'$, and we will see this in a specific example. We begin with reviewing the procedure for performing T duality in the presence of $\alpha'$ corrections in section \ref{SecSubTdualDFT} and commenting on the abelian limit in the end of that subsection. Then in section \ref{SecSubTdualS3} we will perform a non--abelian T duality along a sphere and analyze the resulting quantum corrections.

\subsection{T duality in the DFT prescription}
\label{SecSubTdualDFT}

We begin with a brief overview of the abelian and non--abelian duality procedures in the framework of Double Field Theory. To establish the notation, we begin with the well--known results in the supergravity approximation \cite{OldNATD1w1,OldNATD1w2,OldNATD1w3,OldNATD1w4,1012,1104w1,1104w2,
1104w3,NATD101} before discussing quantum corrections in the end of this subsection. 

Let us consider a non-linear $\sigma$--model on a manifold describing a Lie group $g$. To describe the groups and cosets in a uniform way, we will make only the symmetry under the left action of $G$ manifest, and the most general action with this property can be written as 
\begin{equation}\label{SigmaE}
S=\int d^2\sigma \left[E_{ij}L^i_+L^j_-+Q_{\mu\nu}\d_+Y^\mu \d_-Y^\nu+Q_{\mu i}\d_+Y^\mu L_-^i+
Q_{i\mu }L_+^i \d_-Y^\mu\right]\,.
\end{equation}
Here and below
\bea\label{LeftInvForm}
L^i_\mu=-i\operatorname{Tr}(t^ig^{-1}\partial_\mu g)\quad \mbox{and}\quad L^i_\pm=L^i_\mu \partial_\pm X^\mu
\eea
denote the left invariant Maurer-Cartan forms in the target space and on the worldsheet. By construction, action (\ref{SigmaE}) is invariant under the left action $g\rightarrow h_L g$ for any matrix $E_{ij}$, but for $E_{ij}=E\delta_{ij}$ and $Q_{\mu i}=Q_{i\mu}=0$, i.e., for a sigma model on a group, the symmetry is enhanced to 
$G_L\times G_R$. To dualize the action (\ref{SigmaE}) along $G_L$, one  gauges this isometry, introduces Lagrange multipliers to ensure that the gauge potential has vanishing field strength, and integrates out 
the gauge fields. This leads to the final answer \cite{NATD101}:
\bea\label{basic dual}
S_{\text{dual }}=\int d^2\sigma \left[(\partial_+v_i+\d_+ Y^\mu Q_{\mu i})(M^{-1})^{ij}
(\partial_-v_j-Q_{j\nu}\d_-Y^\nu)+Q_{\mu\nu}\d_+Y^\mu \d_-Y^\nu\right]\,.
\eea
Matrices $M_{ij}$ and $f_{ij}$ are defined in terms of $E_{ij}$, the Lagrange multipliers (dual coordinates) $v_i$, and structure constants $f_{ij}{}^k$  of the group $G$ as,
\begin{equation}\label{dualMij}
M_{ij}=(E_{ij}+f_{ij}), \quad f_{ij}\equiv f_{ij}{}^kv_k\,.
\end{equation}
The metric and the Kalb--Ramond field of the dual theory are encoded in the matrix $M$, and the new dilaton is given by
\begin{equation}\label{basic dilaton}
\phi_{\text{dual}}=\phi-\frac{1}{2} \ln({\operatorname{det}M})\,.
\end{equation}
Expressions (\ref{basic dual}) and (\ref{basic dilaton}) give all NS--NS fields of the dual geometry.

To write the $\alpha'$ corrections to the dual geometry (\ref{basic dual}), authors of \cite{2007.07902} looked at two sets of frames corresponding to the metric (\ref{basic dual}):
\bea\label{DualFrames}
&&{\hat e}^a_+=-{\kappa^a}_j(M^{-1})^{ij}(dv_i+d Y^\mu Q_{\mu i})+{\la^a}_\mu dY^\mu,\nn
&&{\hat e}^a_-={\kappa^a}_i(M^{-1})^{ij}(dv_j- Q_{j\mu}d Y^\mu)+{\la^a}_\mu dY^\mu\,.
\eea
Here matrices $\kappa$ and $\lambda$ are defined by the frames for the original action 
(\ref{SigmaE}):
\bea
e^a=\tensor{\kappa}{^a_i}L^i+\tensor{\lambda}{^a_\mu}dY^\mu,\quad e^A=\tensor{e}{^A_\mu}dY^\mu\,,
\eea
The frames (\ref{DualFrames}) are related by a Lorentz transformation parameterized by matrix $\Lambda$
\begin{equation}\label{lambda group}
\Lambda_i{}^j=-(\kappa^{-T}MM^{-T}\kappa^T)_i{}^j \,. 
\end{equation}
As shown in \cite{2007.07902}, $\alpha'$ corrections to the action (\ref{basic dual}) can be summarized by a simple 
shift in the matrix
\bea
\tilde{M}= \tilde{G}-\tilde{B}\,.
\eea
The result reads
\bea\label{Mtilde}
\Delta \tilde{M}=b\Delta_{{\Lambda}}^{(+)}\tilde{M}
+\alpha^\prime \Delta \tilde{M}^{(1)}\,,
\eea
where $\Delta \tilde{M}^{(1)}$ comes from expanding the original matrix $M$ to the first order in $\alpha'$ in the supergravity relations between $M$ and ${\tilde M}$. This contribution vanishes if the original background is not corrected in the first order. Parameter $b$ in (\ref{Mtilde}) is the same as in (\ref{rpm action}). In particular, $b=-\alpha'$ for the bosonic string and $b=0$ in the heterotc case. The first term in (\ref{Mtilde}) can be written in terms of the matrix (\ref{lambda group}):
\bea\label{MtildePM}
\Delta_{\tilde{\Lambda}}^{(\pm)}\tilde{M}_{mn}&=&\frac{1}{2}\operatorname{tr}\big[\partial_m{\Lambda}{\Lambda}^{-1}\tilde{\omega}^{(\pm)}_n\big]
+\frac{1}{4}\operatorname{tr}\big[\partial_m{\Lambda}{\Lambda}^{-1}\partial_n{\Lambda}{\Lambda}^{-1}\big]-
{B}^{\text{WZW},({\Lambda})}_{mn}\,,\\
d{B}^{\text{WZW},({\Lambda})}&=&-\frac{1}{12}\mbox{tr}\left[d{\Lambda}{\Lambda}^{-1}\wedge d{\Lambda}{\Lambda}^{-1}\wedge d{\Lambda}{\Lambda}^{-1}\right]\,.\nonumber
\eea
These expressions were found in the DFT scheme, and to convert them to the Bergshoeff--de Roo scheme \cite{BdR-1w1,BdR-1w2}, one can use the following relations \cite{2007.07902}
\begin{equation}
\begin{aligned}
\tilde{M}_{mn}^{(BR)}&=\tilde{M}_{mn}+\frac{a}{4}\tilde{\omega}^{(-)}_{ma}{}^b\tilde{\omega}^{(-)}_{nb}{}^a    +\frac{b}{4}\tilde{\omega}^{(+)}_{ma}{}^b\tilde{\omega}^{(+)}_{nb}{}^a,\\
\tilde{\phi}^{(BR)}&=\tilde{\phi}-\frac{a}{4}\tilde{\omega}^{(-)}_{mab}\tilde{\omega}^{(-)mab}    -\frac{b}{4}\tilde{\omega}^{(+)}_{mab}\tilde{\omega}^{(+)mab}\,.
\end{aligned}    
\end{equation}
In this article we are focusing on the DFT prescription.

The corrections (\ref{MtildePM}) can be nontrivial even in the abelian case, where they were discovered in \cite{9705193w1,9705193w2,9705193w3}, but interestingly they vanish for the chiral null model discussed in section \ref{SecNSsoln}. Specifically, T duality interchanges functions $F^{-1}$ and $K$ in the system (\ref{CNMstart}), and in section \ref{SecSubF1} we demonstrated that both functions lead to exact solutions of the action (\ref{rpm action})\footnote{The fact that the chiral null model is exact to all orders in $\alpha'$ in a particular prescription has been known for a long time \cite{CNMw1,CNMw2}, but in section section \ref{SecSubF1} we demonstrated this for the DFT prescription for which  the formulae (\ref{Mtilde}) and (\ref{MtildePM}) are written.}. Therefore the corrections (\ref{MtildePM}) must vanish in this case, and a direct calculation shows that they indeed do. In the next subsection we will consider a more interesting example of T duality where these corrections give a nontrivial contribution.

\subsection{Example of NATD: duality along $S^3$}
\label{SecSubTdualS3}

The duality rules reviewed in the previous subsection are applicable when the elements $g$ in (\ref{SigmaE})--
(\ref{LeftInvForm}) describe a group manifold. The simplest non--abelian group is $SU(2)$, its group manifold is $S^3$, and in this subsection we will explore quantum corrections to a T duality along this space. 

Although we have encountered three--dimensional spheres in section \ref{SecRmin}, unfortunately the duality construction outlined above is not applicable to spaces (\ref{NSfinProd}): while the metric and the $H$ field are 
group--invariant, the $B$ field is not. On a technical level, one encounters an obstruction in writing the sigma model for (\ref{NSfinProd}) in the form (\ref{SigmaE})--(\ref{LeftInvForm}). Since it is not possible to define a group--invariant 
$B$ field for (\ref{NSfinProd}), in this subsection we will focus on a geometries that has a form
\bea\label{S3geomStart}
ds^2=ds_\perp^2+\rho^2 d\Omega_3^2,\quad B=B_\perp,\quad e^{2\phi}=e^{2\phi_\perp}\,,
\eea
so that the Kalb--Ramond field has no components along the sphere directions. We will further assume that the geometry solves equations of motion coming from the action (\ref{rpm action}) with trivial $\alpha'$ corrections. Examples of such situations are flat space and the chiral null model discussed in section \ref{SecSubF1}. In the latter case we find
\bea\label{CNMmetrCopy}
ds^2&=&F[du dv+K du^2]+dx_adx_a+d\rho^2+\rho^2d\Omega_3^2,\nn
B&=&F\,du\wedge dv,\quad \phi=\phi_0+\frac{1}{2}\ln F.
\eea
Performing the non--abelian T duality along the sphere directions in (\ref{S3geomStart}) and using the results of \cite{2007.07902,2007.07897,2007.09494} reviewed in the previous subsection to evaluate the $\alpha^\prime$ corrections, we arrive at the final answer:
\bea\label{S3geomFinish}
d\tilde{s}^2&=&ds_\perp^2+\frac{1}{4}\left[\frac{dr^2}{\rho^2}+\frac{r^2\rho^2}{\rho^4+r^2}d\Omega_2^2\right]
-b\frac{(3\rho^4+r^2)}{(\rho^4+r^2)^2}\left[{dr^2}+r^2d\Omega_2^2\right],\nn
\tilde{B}&=&-\frac{r^3}{4[\rho^4+r^2]}\bigg[1-b\frac{8\rho^2}{(\rho^4+r^2)}\bigg]\operatorname{Vol}(S^2),\\
e^{2\tilde{\phi}}&=&e^{2\phi_\perp}\left[\frac{64}{\rho^2(\rho^4+r^2)}\right]
\left[1-2b \frac{(3\rho^4+r^2)(3\rho^4+2r^2)}{\rho^2(\rho^4+r^2)^2}\right]\,.\nonumber
\eea
Recall that $b=-\alpha'$ for the bosonic string, so the classical geometry receives nontrivial $\alpha'$ corrections, and $b=0$ in the heterotic case.
As expected, no singularities are generated away from the points where $\rho=0$: at the potentially dangerous locations where $r$ becomes small, one finds
\bea
d\tilde{s}^2\simeq ds_\perp^2+\left[\frac{1}{4\rho^2}-\frac{3b}{\rho^4}\right]\left[{dr^2}+{r^2}d\Omega_2^2\right],\quad 
\tilde{B}\simeq-\frac{r^3}{4\rho^4}\bigg[1-\frac{8b}{\rho^2}\bigg]\operatorname{Vol}(S^2).\nonumber
\eea
We performed the duality along $S^3$ since it can be viewed as a group manifold. To carry out this operation for other spheres, one needs to extend the procedure described in the previous subsection to cosets, and this is an interesting open problem. 

\section{Discussion}

In this article we explored quantum corrections to several classes of supergravity solutions. Starting with the action (\ref{rpm action}) obtained using the framework of the Double Field Theory, we constructed all classical geometries that don't receive quantum corrections due to vanishing of every single term in (\ref{rpm action}). We also analyzed several examples of solutions for which quantum corrections vanish due to cancellations, as well as some geometries with nontrivial corrections. Finally, we presented an example of quantum corrections generated by the non--abelian T duality.

\bigskip

Let us briefly summarize our results. The action (\ref{rpm action}) derived in the DFT prescription is naturally written in terms of modified curvatures $R^{(\pm)}$, and all $\alpha'$ corrections disappear when such curvatures vanish. In section \ref{SecRmin} we showed that such term--by--term disappearance of corrections implies that the geometry must be a product of group manifolds and flat space with some additional NS--NS fluxes. Furthermore, in the analogy with Riemann--flat spaces that admit coordinate frames with flat spin connection, vanishing of $R^{(\pm)}$ implies existence of the unique frames (\ref{new frames}) with vanishing twisted spin connections, and the standard spin connections in these frames turned out to be constant. Since from the perspective of string theory, the most interesting group manifold is a three--dimensional sphere, this raised an interesting possibility for special frame on $n$--spheres: while twisted connections in this case are not very useful, it might be possible to define frames with constant spin connections. In section \ref{SecFrameSn} we constructed such frames for all spheres and showed that they were unique up to constant Lorentz rotation. In the two--dimensional case, we also demonstrated that the requirement of constant spin connections implied that the geometry must be a sphere, a hyperboloid, or a flat space. It would be interesting if this result can be extended to higher dimensions, i.e., if constant connections imply that the geometry must be a product of spheres, hyperboloids, and flat spaces. Our results on connections and frames are summarized in Table \ref{Table1}. 

While vanishing of the twisted curvature $R^{(\pm)}$ guarantees that quantum corrections disappear term--by--term, some more general backgrounds are exact due to various cancellations. In section \ref{SecNSsoln} we considered one of such examples, the chiral null model, which is known to be exact in some prescription for computing quantum corrections \cite{CNMw1,CNMw2}. We showed that such corrections cancel in the DFT prescription as well, even though the individual components of $R^{(\pm)}$ are highly nontrivial. In the same section we also analyzed the geometries produced by NS5 branes and computed the leading quantum corrections, which turned out to be nontrivial. In the near--horizon limit, these geometries reduce to the exact solution describing three--sphere with NS--NS fluxes. As another example of geometry with nontrivial quantum corrections, we analyzed small gravitational waves in section \ref{SecGW}. In the supergravity approximation, the only propagating degrees of freedom are massless photons, but quantum corrections introduce additional modes whose mass goes to infinity as $\alpha'$ is sent to zero. We explicitly constructed such solutions in section \ref{SecGW}.

Finally, in section \ref{SecTdual} we addressed the question of T duality. After briefly reviewing the duality procedure in both abelian and non--abelian cases, including $\alpha'$ corrections, we applied this operation to geometries containing a three--dimensional sphere. In contrast to the supergravity approximation, where one can dualize along any sphere by viewing it as a coset, currently the procedure for computing $\alpha'$ corrections is available only in the group case, making $S^3$ the unique physically relevant manifold for performing the duality. It would be very interesting to extend the procedure for duality along cosets beyond supergravity approximation, allowing one to dualize along spheres of various dimensions that appear in various string backgrounds.

Since in this work we have been studying solutions associated with the DFT--inspired action 
(\ref{DFT action}), (\ref{rpm action}), which contains only NS--NS fields in bosonic and heterotic strings, unfortunately the most interesting backgrounds of type IIB theory, which often contain Ramond--Ramond fluxes are beyond the scope of our investigation. At the supergravity level, enormous progress has been made in studying such backgrounds in the framework of DFT \cite{DFTclassW1,DFTclassW2,DFTclassW3,DFTclassW4,DFTclassW5,DFTclassW6,DFTclassW7,
DFTclassW8,
DFTclassW9,DFTclassW10,DFTclassW11,DFTclassW12,DFTclassW13}, but these geometries receive quantum corrections only at $O(\alpha'^3)$ level and higher. It would be very interesting to extend the actions (\ref{DFT action}), (\ref{rpm action}) to that level and to incorporate the RR fields into such extensions.

\section*{Acknowledgements}

This work was supported in part by the DOE grant DE-SC0015535.

\appendix
\section{Conventions}\label{conventions}
In this appendix we summarize the notation used throughout the paper. We mostly follow the conventions used in \cite{1507}.

Starting with a background described by a metric $g_{\mu\nu}$ and a Kalb--Ramond field $B_{\mu\nu}$, we introduce frames $e_{\mu}{}^a$ and partial projections $H_{\mu a}{}^b$ of the three--form field strength:
\bea\label{DefH app}
H_{\mu\nu\rho}=\partial_\mu B_{\nu\rho}+\partial_\nu B_{\rho\mu}+\partial_\rho B_{\mu\nu},\quad H_{\mu a}{}^b=H_{\mu\nu\rho}e_a{}^\nu g^{\rho\sigma}e_\sigma{}^b.
\eea
Here $\mu,\nu,\ldots$ are the curved spacetime indices and $a,b,\ldots$ denote the flat directions. The spin connection $\omega_{\mu}{}^{ab}$ are defined by 
\bea\label{OmegaForm}
de^a+{\omega_\mu}^{ab} dx^\mu\wedge e_b=0,
\eea
or, more explicitly, by
\begin{equation}
\omega_{\mu}{}^{ab}=\frac{1}{2}\big[e^{a \mu}(\partial_\mu e_{\nu}{}^b-\partial_\nu e_{\mu}{}^b)-e^{b\nu}(\partial_\mu e_{\nu}{}^a-\partial_\nu e_{\mu}{}^a)-e^{a\rho}e^{b\sigma}(\partial_\rho e_{\sigma c}-\partial_\sigma e_{\rho c})e_{\mu}{}^c\big]\,,
\end{equation}
and they can be used to evaluate the partially projected Riemann tensor: 
\begin{equation}\label{R tensor}
R_{\mu\nu ab}(\omega)=\partial_\mu\omega_\nu{}^{ab}-\partial_\nu\omega_{\mu}{}^{ab}+\omega_{\mu}{}^{ad}\eta_{cd}\omega_\nu{}^{cb}    -\omega_{\nu}{}^{ad}\eta_{cd}\omega_\mu{}^{cb}\,.
\end{equation}
The Riemann tensor with spacetime indices can be expressed in terms of the Christoffel symbols via the standard relation:
\bea\label{R in terms of Gamma}
R^\rho{}_{\sigma\mu\nu}=-R_{\mu\nu a}{}^be_b{}^\rho e_\sigma{}^a:\quad 
R^\rho{}_{\sigma\mu\nu}=\partial_\mu\Gamma^\rho_{\nu\sigma}-\partial_\nu\Gamma^\rho_{\mu\sigma}+\Gamma^\rho_{\mu\delta}\Gamma^\delta_{\nu\sigma}-\Gamma^\rho_{\nu\delta}\Gamma^\delta_{\mu\sigma}.
\eea
We will also use the Chern-Simons 3-form constructed from the spin connections as
\bea\label{CS 3-from}
\Omega_{\mu\nu\rho}(\omega)=\omega_{[\mu a}{}^b\partial_\nu\omega_{\rho]b}{}^a+\frac{2}{3}\omega_{[\mu a}{}^b\omega_{\nu b}{}^c\omega_{\rho]c}{}^a~.
\eea
While the supergravity action can be nicely written in terms of the Riemann tensor and the three--form fields strength $H$, the alpha prime corrections are naturally expressed in terms of modified (``twisted'') 
spin-connection and curvature which include torsion and which mix the metric and the Kalb--Ramond field. The modified spin connections are defined by
\bea\label{DefOmegaPM}
\omega^{(\pm)}_{\mu a}{}^b=\omega_{\mu a}{}^b\pm\frac{1}{2}H_{\mu a}{}^b\,,
\eea
and the modified Chern-Simons 3-form and Riemann tensor are defined by the counterparts of 
equations (\ref{CS 3-from}) and (\ref{R tensor}) with $\omega^{(\pm)}_{\mu a}{}^b$ instead of $\omega_{\mu a}{}^b$. The modified Riemann tensor can also be written as 
\bea
R^{(\pm)}_{\mu\nu a b}=R_{\mu\nu a b}(\omega)\pm\mathcal{D}_{[\mu}H_{\nu]ab}-\frac{1}{2}H_{[\mu a }{}^cH_{\nu]bc}\,,
\eea
where the Lorentz covariant derivative acts as
\bea
\mathcal{D}_\mu T_a{}^b=\partial_\mu T_a{}^b+\omega_{\mu a }{}^cT_c{}^b-\omega_{\mu c}{}^bT_a{}^c~.
\eea
This concludes our brief summary of formulas from differential geometry, let us now make some comments about effective actions for strings.

At the supergravity level, the equations of motion for the NS--NS fields are the same for all string theories, and they are given by\footnote{We assume that all other fields, such as Ramond--Ramond fluxes in type II or gauge fields in the heterotc strings are turned off. In the presence of such fields, even equations for the metric, the dilaton, and the Kalb--Ramond field depend on the theory.}
\begin{equation}\label{eom lading order}
  \begin{aligned}
&R_{\mu\nu}+2\nabla_\mu\nabla_{\nu}\phi-\frac{1}{4}H_{\mu\kappa\sigma}H_{\nu}{}^{\kappa\sigma}=0,\\
&-\frac{1}{2}\nabla^\kappa H_{\kappa\mu\nu}+\nabla^\kappa\phi H_{\kappa\mu\nu}=0,\\
&-\frac{1}{2}\nabla^2\phi+\nabla_\kappa\phi\nabla^\kappa\phi-\frac{1}{24}H_{\kappa\mu\nu}H^{\kappa\mu\nu}=0.
  \end{aligned}  
\end{equation}
In contrast to this universality, quantum corrections depend not only on a theory, but also on a scheme used in computing string amplitudes, and in the heterotc case, $(a,b)=(-\alpha^\prime,0)$ in the notation of (\ref{L1}), we follow the conventions of \cite{1507} based on the Bergshoeff-de Roo scheme \cite{BdR-1w1,BdR-1w2}:
\begin{equation}\label{omega minus}
\begin{aligned}
S=\int dx\sqrt{-g}e^{-2\phi}\bigg[&R-4\nabla_\mu\phi\nabla^\mu\phi+4\nabla^\mu\nabla_\mu\phi-\frac{1}{12}\tilde{H}_{\mu\nu\rho}\tilde{H}^{\mu\nu\rho}+\frac{a}{8}R^{(-)}_{\mu\nu a}{}^bR^{(-)\mu\nu}{}_{b}{}^a\bigg],  \\
\tilde{H}_{\mu\nu\rho}&=H_{\mu\nu\rho}-\frac{3a}{2}\Omega^{(-)}_{\mu\nu\rho},\quad
\omega^{(-)}_{\mu a}{}^b=\omega_{\mu a}{}^b-\frac{1}{2}H_{\mu a}{}^b.
\end{aligned}    
\end{equation}
It is instructive to compare this with the results of \cite{1608}, where the effective action for the heterotic string was written as
\bea\label{tempSnew}
\hskip -0.7cm
S=\int dx \sqrt{-g}e^{-2\phi}\bigg[&R(\omega)-\frac{1}{3}\tilde{H}_{\mu\nu \rho}\tilde{H}^{\mu\nu \rho}+4\partial_\mu\phi\partial^\mu\phi+\alpha^\prime R_{\mu\nu ab}(\omega_{+})R^{\mu\nu ab}(\omega_{+})\bigg]\,,
\eea
with
\bea
\omega_{\pm \mu}{}^{ab}&=\omega_{\mu}{}^{ab}\pm \tilde{H}_{\mu}{}^{ab},\quad
\tilde{H}_{\mu\nu\rho}=H_{\mu\nu\rho}-\alpha^\prime\Omega^{(+)}_{\mu\nu\rho}\,.
\eea
The difference between (\ref{omega minus}) and (\ref{tempSnew}) can be traced back to a somewhat non--standard definition of $H_{\mu\nu\rho}$ used in \cite{1608}:
\bea
H_{\mu\nu\rho}=-\frac{1}{2}(\partial_\mu B_{\nu\rho}+\partial_\nu B_{\rho\mu}+\partial_\rho B_{\mu\nu}).\nonumber
\eea
The difference between the last relation and (\ref{DefH app}) is responsible for mapping 
$\omega^{(-)}_{\mu a}{}^b$ from (\ref{omega minus}) into
$\omega^{(+)}_{\mu a}{}^b$ in the conventions of \cite{1608}.

\section{Perturbative analysis of the condition $R^{(-)}=0$}
\label{SecAppRmPert}

In this appendix we provide some technical details leading to justification of equation (\ref{ZconstrEqn}) as the most general condition for the geometries solving the constraint (\ref{RminZero}). Then we will show that in the most interesting case of $d\le 6$ dimensional manifold with fluxes, the Jacobi identity (\ref{ZconstrEqn}) leads to a product of three--spheres  (\ref{NSfinProd}), while more generally we recover the Cartan--Schouten result (\ref{NSfinProdG}).

As the first step in analyzing the equation (\ref{RminZero}),
\bea\label{RminZeroAp}
R^{(-)}_{iklm}=0,
\eea
we look at an expansion near some generic point. By choosing convenient coordinates and imposing a gauge on the two--form, we can expand the fields as
\bea\label{gBexpandApp}
g_{mn}=\delta_{mn}+\eps h_{mn}+\eps^2 h^{(2)}_{mn}+\dots,\quad
B_{mn}=\eps b_{mn}+\eps^2 b^{(2)}_{mn}+\dots,
\eea
and treat $\eps$ as a small parameter. Substitution of these expansions into the expression (\ref{RpmDef}) for $R^{(-)}$ gives
\bea
R^{(-)}_{iklm}=\frac{\eps}{2}\left[\d_k\d_l h^{(-)}_{im}-\d_i\d_lh^{(-)}_{km}-\d_k\d_m h^{(-)}_{il}+
\d_i\d_m h^{(-)}_{kl}\right]+O(\eps^2),\quad h^{(-)}_{mn}=h_{mn}-b_{mn}\,.\nonumber
\eea
In particular, the metric in the Kalb--Ramond field can be decoupled by taking two linear combinations:
\bea\label{RplusDecplApp}
&&R^{(-)}_{iklm}+R^{(-)}_{lmik}=\eps\left[\d_k\d_l h_{im}-\d_i\d_l h_{km}-\d_k\d_m h_{il}+
\d_i\d_m h_{kl}\right]+O(\eps^2),\nn
&&R^{(-)}_{iklm}-R^{(-)}_{lmik}=-\eps\left[\d_k\d_l b_{im}-\d_k\d_m b_{il}+\d_i\d_m b_{kl}-\d_i\d_lb_{km}\right]+O(\eps^2).
\eea
Equation (\ref{RminZeroAp}) implies that both lines must vanish. The first line is the first term in the expansion of the Riemann tensor around flat space, and it vanishes if and only if $h_{mn}$ is a pure gauge. Therefore by performing an $\eps$--dependent diffeomorphism, we can set $h_{mn}=0$ in (\ref{gBexpandApp}). Imposing equations (\ref{RminZeroAp}) in the second line of (\ref{RplusDecplApp}), we arrive at relations
\bea\label{Heqn1ord}
\d_k c_{lim}-\d_ic_{klm}=0,\quad \mbox{where}\quad c_{lim}=\d_l b_{im}+\d_m b_{li}+\d_i b_{ml}\,.
\eea
As expected from equation (\ref{gBexpandApp}), tensor $c_{lim}$ is the leading contribution to the $H$--field. The 
$(1213)$ component of (\ref{Heqn1ord}) gives $\d_1 c_{213}=0$, and similar relations imply that $c_{ijk}$ does not depend on the coordinates $(x_i,x_j,x_k)$. In the three--dimensional case this implies that $c_{123}=\mbox{const}$. If $d>3$, then the $(1234)$ component of (\ref{Heqn1ord}) gives
\bea
\d_1 c_{234}-\d_2 c_{134}=0\quad\Rightarrow\quad \d^2_1 c_{234}=0.
\eea
Application of a similar logic to other combinations of indices leads to the conclusion that 
all components of the $c$ field are at most linear in coordinates. Constant $H$ field solves all equations, so we focus on linear contributions in the four dimensional subspace:
\bea\label{tempLinH}
c_{123}=a_4 x_4,\quad c_{234}=a_1 x_1,\quad c_{134}=a_2 x_2,\quad c_{124}=a_3 x_3\,.
\eea
Here coefficients $a_k$ may depend on $(x_5,\dots x_d)$, but not on $(x_1,\dots,x_4)$. 
Substitution in the second equation in (\ref{RplusDecplApp}) gives the following vanishing components:
\bea
&&E_{1234}=a_1-a_2,\ E_{1324}=-a_1-a_3,\
E_{1423}=a_1-a_4,\ E_{2314}=-a_2+a_3,\nn
&&E_{2413}=a_2+a_4,\ E_{3412}=a_3-a_4\,.
\eea
Consistency implies that $a_k=0$, i.e., that the $H$ field for the system (\ref{tempLinH}) vanishes. An absence of the nontrivial solution (\ref{tempLinH}) implies that, in the leading order in $\eps$, the  $H$ field must be constant. 

In the second order in $\eps$, we find
\bea
\frac{\eps^2}{2}\left[\d_k\d_l h^{(-,2)}_{im}-\d_i\d_lh^{(-,2)}_{km}-
\d_k\d_m h^{(-,2)}_{il}+
\d_i\d_m h^{(-,2)}_{kl}\right]=\frac{\eps^2}{2}\left[c_{ils}c_{kms}-c_{kls}c_{ims}\right].
\eea
Separating $h^{(-)}$ into the metric and the B field as before, we arrive at the homogeneous equation for the $B$ field, so it is a pure gauge. For the metric, we find a linear combination of a pure gauge and a special solution in 4d:
\bea
h^{(2)}_{mn}dx^m dx^n&=&\frac{(c_{123})^2}{4}(dy_1 dy_2+dy_1 dy_3+dy_2 dy_3)+
\frac{(c_{124})^2}{4}(dy_1 dy_2+dy_1 dy_4+dy_2 dy_4)\nn
&&+\frac{c_{123}c_{124}}{2}(dy_1+dy_2)(x_3dx_4+x_4 dx_3).
\eea
Here we defined convenient combinations $y_k=(x_k)^2$. The last relation implies that there are no obstructions in developing perturbation theory starting from arbitrary constant $c_{ijk}$ in three and four dimensions.

In five dimensions, we can try to switch on $c_{123}$ and $c_{145}$. It turns out that there are no solutions in the second order unless the cross term vanishes: $c_{123}c_{145}=0$. To be consistent with symmetries, an invariant way of eliminating such terms must have the form 
\bea
\left[\alpha_1 c_{iks}c_{lms}+perm\right]+\left[\beta_1\delta_{il}c_{tks}c_{tms}+perm\right]=0
\eea
with some undetermined constant $(\alpha_1,\alpha_2,\alpha_3,\beta_1,\dots\beta_6)$. Imposing this constraint on the $(c_{123},c_{124})$ pair in 5d, which should satisfy the constraint, we find the unique solution
\bea\label{ZconsApp}
Z_{ijkl}=c_{ijs}c_{kls}+c_{iks}c_{ljs}+c_{ils}c_{jks}=0.
\eea
As expected, this constraint is violated by the $(c_{123},c_{145})$ pair. Since the constraint (\ref{ZconsApp}) is antisymmetric in any pair of indices, it gives at most $d(d-1)(d-2)(d-3)/24$ conditions, but there is an additional reduction due to an identity
\bea
c_{ija}Z_{ijbc}=c_{ija}c_{ijs}c_{bcs}+c_{ija}c_{sib}c_{jsc}-c_{ija}c_{sic}c_{jsb}=0.
\eea
Let us now use rotations to reduce the number of components of $c_{ijk}$
\begin{enumerate}[(a)]
\item $d=3$: there is only one component, $c_{123}$, and it solves the constraint.
\item $d=4$: three--form can be dualized into a vector, which can be rotated to $V_1=V_2=V_3=0$. This again leads to $c_{123}$.
\item $d=5$: three--form can be dualized into a two--form, and the latter can be rotated into $(W_{23},W_{45})$. This translates into $(c_{123},c_{145})$, and the constraint forces the product $c_{123}c_{145}$ to vanish. 
\item $d=6$: in the 5d subspace we choose $(W_{23},W_{45})$, then subsequent rotations in $(23)$, $(45)$ planes can be used to simplify $c_{6ij}$. The surviving components are
\bea
c_{123},\ c_{145},\ c_{624},\ c_{635},\ c_{623},\ c_{645},\ c_{61i}\,.
\eea
In terms of counting, we start with 20 components and remove 15 rotations, so there are at most five independent components. Assuming that $c_{123}$ is not equal to zero, we find four solutions:
\begin{enumerate}[A.]
\item $(c_{123},c_{126},c_{136},c_{236})$ with four arbitrary coefficients. Rotation in the $(2,3)$ plane sets $c_{136}=0$, then rotation in the $(1,6)$ plane sets $c_{236}=0$. This leaves only two parameters $(c_{123},c_{126})$, and rotation in the $(3,6)$ plane leads to only one nontrivial component $c_{123}$.
\item $(c_{123},c_{456})$ with two arbitrary coefficients. This is a generalization of A.
\item Three--parameter solution: $c_{145}=-c_{236}c_{456}/c_{123}$. Rotation in the $(1,6)$ plane can be used to set $c_{236}=0$ leading to option B. 
\item Three--parameter solution with $c_{123}=c_{145}$, $c_{624}=c_{635}=\sqrt{(c_{123})^2+c_{236}c_{456}}$, up to reflection of axes. 
\end{enumerate}
In the cases B--D there are two distinct eigenvalues of the matrix $M_{ab}=c_{aij}c_{bij}$. By performing a rotation, this matrix can be put in the form
\bea
M_{ab}=\la_1 P^{(123)}_{ab}+\la_2 P^{(456)}_{ab}\,,
\eea
where $P_{ab}$ denote projections into three--dimensional subspaces. Then rotations in two 3d spaces can be used to keep only 
\bea
c_{123},\ c_{456},\ c_{234},\ c_{1ij},\ c_{2ij},\ c_{3ij}:\ (ij)=\{456\}\,.
\eea
Rotations in $(23)$ and $(56)$ plane lead to
\bea
c_{123},\ c_{456},\ c_{234},\ c_{1ij},\ c_{245},\ c_{346},\ c_{256},\ c_{356}\,.
\eea
Solutions of the constraint are
\begin{enumerate}[A.]
\item $(c_{123},c_{456})$.
\item $c_{456}=c_{123}$,\ $c_{156}=-c_{234}$. Rotation in $(14)$ plane leads to A with $c_{456}=c_{123}$.
\item $c_{245}=c_{123}=c_{156}=c_{346}$.
\item $c_{245}=c_{123}=a$, $c_{146}=-c_{356}=b$, $c_{156}=c_{346}=\sqrt{a^2-b^2}$. Rotation in $(45)$ direction can be used to set $b=0$. 
\item $c_{346}=c_{123}=a$, $c_{145}=-c_{256}=b$, $c_{156}=c_{245}=\sqrt{a^2-b^2}$. Rotation in $(46)$ direction can be used to set $b=0$. Then options D and E unify into C.
\end{enumerate}
\end{enumerate}
Detailed analysis of the constraints (\ref{RminZero}) for $d>6$ is beyond the scope of this article. We just observe that the relation (\ref{RminZero}) can be interpreted as the Jacobi identity if coefficients $c_{abc}$ are viewed as structure constants of a Lie algebra: 
\bea\label{CabcLam}
c_{abc}=\la f_{abc},\quad [T_a,T_b]={f_{ab}}^cT_c
\eea
Then introducing the frames for the corresponding Lie group
\bea
e^a=\mbox{tr}(g^{-1}dg T^a),
\eea
we define the Kalb--Ramond three--form by
\bea
H=\frac{1}{6}\la f_{abc}e^a\wedge e^b\wedge e^c
\eea
Using the Maurer--Cartan relations
\bea\label{Maurer}
de^a=-e^b\wedge e^c\, \mbox{tr}(T_b T_c T^a)=-\frac{1}{2}{f_{bc}}^a\, e^b\wedge e^c
\eea
and the Jacobi identity, one can verify that $dH=0$. 
Comparing (\ref{Maurer}) with the definition (\ref{OmegaForm}),
one extracts the frame compoments of the spin connection:
\bea
{\omega_c}^{ab}=\frac{1}{2}{f_{c}}^{ab}.
\eea
Recalling the definition (\ref{DefOmegaPM}) of the modified spin connection, we conclude that the geometry 
\bea
ds^2=e^a e^a,\quad H=\frac{1}{6}f_{abc}e^a\wedge e^b\wedge e^c
\eea
has vanishing $\omega^{(-)b}_{\mu a}$ and therefore vanishing $R^{(-)}_{\mu\nu ab}$. This is just a reformulation of the well-known Cartan--Schouten result \cite{Cartan1,Cartan2}. Putting back the rescaling by factor $\la$ from (\ref{CabcLam}) for each group, we arrive at the final answer (\ref{NSfinProdG}).

\section{Special frames in two dimensions}
\label{AppFrame2d}

In section \ref{SecFrameSn} we constructed special frames for various spheres. Specifically, we showed that on every sphere $S^n$, one can define frames which lead to constant spin connections $\omega^{ab}_c$, and the choice of such frames is unique up to a rotation. This brings an interesting inverse question: does existence of frames with constant $\omega^{ab}_c$ imply that the geometry is a product of spheres, hyperboloids and flat space? In other words, can the arrow in the third line of Table \ref{Table1} be reversed? For the first two lines, the answer is affirmative: existence of frames with vanishing connections (twisted connections) implies that the space is flat (a product of flat subspaces and three--spheres)\footnote{For the vanishing connections this result is well known, Vanishing of twisted connections implies vanishing of $R^{(-)}_{\mu\nu\la\sigma}$, then the discussion presented in section \ref{SecSubRmDerv} implies that the space is a product of flat subspaces and three--spheres.}. For the third line we were not able to arrive at a definitive conclusion, but in this appendix we will show that in two dimensions, the arrow can indeed be reversed. In other words, a two--dimensional space constant $\omega^{ab}_c$ must have constant curvature, so it is a sphere, hyperboloid, or flat space. 

\bigskip

In two dimensions, any metric can be written in terms of one function $H$:
\bea\label{May31Metr}
ds^2=2H dx_1dx_2\,.
\eea
Then defining the frames
\bea
e^+=f_1 dx_1+g_1 dx_2,\quad e^-=f_2 dx_2+g_2 dx_1\,,
\eea
we conclude that $f_1 g_2=0$, so without loss of generality, we assume that $g_2=0$, while both $f_1$ and $f_2$ are non--zero. This implies that $g_1=0$ and
\bea
e^+=f_1 dx_1,\quad e^-=\frac{H}{f_1} dx_2.
\eea
 The frame components of the spin connections are
\bea
{e^+}_{+-}=-\frac{1}{H}{\d}_2f_1\equiv iw_1,\quad 
{e^-}_{+-}=\frac{1}{H}\d_1\frac{H}{f_1}\equiv iw_2\,,
\eea
and we require them to be constant. 
If the first component is zero, then $H=H_1(x_1)H_2(x_2)$, and redefining coordinates, we conclude that (\ref{May31Metr}) is flat space. 

If $w_1$ is nonzero, then
\bea
H=-\frac{1}{iw_1}{\d}_2f_1\,,
\eea
and function $f_1$ obeys a differential equation
\bea
\d_1{\d_2}\ln{f_1}=iw_2{\d_2}f_1\,.\nonumber
\eea
Integrating this equation and writing $f_1=e^f$, we find
\bea
\d_1 f-iw_2e^f+h(x_1)=0.
\eea
To proceed we write function $h$ in terms of a new function $g_1(x_1)$ and shift function $f$ as
\bea
h=-\d_1\ln g'_1,\quad f={\tilde f}+\ln g'_1\,.\nonumber
\eea
This leads to the equation
\bea
 \d_1{\tilde f}=iw_2 g_1'({x_1})e^{\tilde f}\,.
\eea
Integrating this relation, we arrive at the final answer:
\bea
f_1=\frac{ig_1'(x_1)}{w_2[g_1(x_1)+g_2(x_2)]},\quad
H=\frac{g_1'g_2'}{w_1w_2[g_1+g_2]^2},\quad f_2=-\frac{ig_2'(x_2)}{w_1[g_1(x_1)+g_2(x_2)]}\,.
\eea
Using reparametrization of $x_1$ and $x_2$, one can choose new coordinates $z=g_1(y_1)$, 
${\bar z}=g_2(y_2)$ obtain the standard metric of the sphere or a hyperboloid:
\bea
ds^2=\frac{2}{w_1w_2(z+{\bar z})}dz d{\bar z}\,,
\eea
and the corresponding frames
\bea
e^+=\frac{idz}{w_2(z+{\bar z})},\quad e^-=-\frac{id{\bar z}}{w_1(z+{\bar z})}\,.
\eea
On the other hand, setting 
\bea
&&g_1=\frac{w_1 z}{1+iw_1 z},\quad g_2=\frac{i}{1+iw_1 {\bar z}},\nn
&&f_1=\frac{w_1}{w_2}\frac{1+iw_1{\bar z}}{1+iw_1 z}\frac{1}{1+w_1^2 z{\bar z}},
\quad f_2=\frac{w_1}{w_2}\frac{1+iw_1 z}{1+iw_1{\bar z}}\frac{1}{1+w_1^2 z{\bar z}}\,,
\eea
we recover the holomorphic frames (\ref{S2holom}). 

\section{Equations of motion for the chiral null model}
\label{SecAppCNM}

In this appendix we analyze equations of motion generated by the DFT Lagrangian with leading $\alpha'$ corrections (\ref{L0})--(\ref{L1}) for the ansatz corresponding to the chiral null model (\ref{CNMmetr}). While the CNM is known to be exact in a certain prescription used in \cite{CNMw1,CNMw2}, it is instructive to verify that no $\alpha'$ corrections are generated in the DFT prescription as well. Our final answers, (\ref{CNMeqnF3}), (\ref{CNMeqnF2}), (\ref{CNMeqnF1}), show that this is indeed the case. Rather that looking at full equations of motion, we will use the minisuperspace approach and evaluate the action as a functional of only three functions appearing in the ansatz (\ref{CNMmetr}). 

\bigskip

To derive the equations of motion for the ansatz (\ref{CNMmetr}),
\bea
ds^2&=F_1[du dv+K du^2]+dx_idx_i,\nn
B&=F_2\,du\wedge dv,\quad \phi=\phi_0+\frac{1}{2}\ln F_3,    
\eea
we begin with evaluating various components of the Riemann tensor. The nontrivial components of the Christoffel symbols are given by
\begin{equation}
\begin{aligned}
&\Gamma^0_{i0}=\frac{1}{2}\bigg[\frac{\partial_iF_1}{F_1}-\partial_iK\bigg],\quad \Gamma^1_{1i}=\frac{1}{2}\bigg[\frac{\partial_iF_1}{F_1}+\partial_iK\bigg],\quad
\Gamma^0_{1i}=\frac{1}{2}(\partial_iK),\\
&\Gamma^i_{00}=\frac{1}{2}[\partial^iF_1-K\partial^iF_1-F_1\partial^iK],\quad
\Gamma^i_{11}=-\frac{1}{2}[\partial^iF_1+K\partial^iF_1+F_1\partial^iK],
\\
&\Gamma^1_{0i}=-\frac{1}{2}(\partial_iK),\quad 
\Gamma^i_{10}=\frac{1}{2}(K\partial^iF_1+F_1\partial^iK),
\end{aligned}    
\end{equation}
and they lead to non-zero components of the Riemann tensor:
\bea
R_{0i0j}&=&\frac{1}{4F_1}\bigg[2F_1{\partial_j\partial_iF_1}-{\partial_iF_1\partial_jF_1}\bigg]-R_{0i1j},\quad R_{0101}=\frac{1}{4}(\partial_iF_1\partial^iF_1),\nn
R_{1i1j}&=&\frac{1}{4F_1}\bigg[-2F_1{\partial_j\partial_iF_1}+{\partial_iF_1\partial_jF_1}\bigg]-R_{0i1j},\\
R_{0i1j}&=&\frac{1}{4F_1}\bigg[2{F_1^2\partial_i\partial_jK}+{F_1\partial_iK\partial_jF_1}+{F_1\partial_jK\partial_iF_1}+{2F_1K\partial_j\partial_iF_1}-{K\partial_iF_1\partial_jF_1}\bigg].
\nonumber
\eea
Computing the Ricci scalar and using the expression for the three--form field strength, 
$H_{i01}=-\partial_iF_2$, we arrive at the Lagrangian (\ref{L0}) in the minisuperspace truncation:
\begin{equation}
\begin{aligned}
L_{(0)}=&\bigg[-\frac{2}{F_3}(\partial^i\partial_iF_1)+\frac{1}{2F_1F_3}(\partial_iF_1)(\partial^iF_1)+\frac{1}{2F_1F_3}(\partial_iF_2)(\partial^iF_2)\\
&+\frac{2F_1}{F_3^2}(\partial_i\partial^iF_3)-\frac{3F_1}{F_3^3}(\partial_iF_3)(\partial^iF_3)+\frac{2}{F_3^2}(\partial_iF_1)(\partial^iF_3)\bigg]
\end{aligned}    
\end{equation}
To evaluate the first $\alpha'$ correction (\ref{L1}), we need to pick frames, and the natural choice is 
\begin{equation}
\begin{aligned}
e_0{}^0=\frac{F_1}{\sqrt{F_1(1+K)}},\quad e_0{}^1=\frac{-F_1K}{\sqrt{F_1(1+K)}},\quad e_1{}^1=\sqrt{F_1(1+K)},\quad e_i{}^j=\delta_i^j.  
\end{aligned}    
\end{equation}
Then evaluation of $L^{(1)}$ leads to the expression (\ref{LagrCNMms}). Variation of
$L^{(0)}+L^{(1)}$ with respect to $F_3$ gives
\begin{equation}\label{CNMeqnF3}
\begin{aligned}
&\frac{1}{2F_1F_3^4}\big[{4}{F_1F_3^2}(\nabla^2 F_1)-
{F_3^2}[(\nabla F_1)^2+(\nabla F_2)^2]-{4F_1^2}{F_3}(\nabla^2 F_3)\\
&\qquad+{6F_1^2}{}(\nabla F_3)^2-\frac{4}{F_3F_1}(\partial_kF_1)(\partial^kF_3)\big]\\
&=\frac{-\alpha^\prime}{4F_1^3F_3^2}\bigg[[(\partial_iF_1)(\partial^iF_2)]^2-\frac{7}{2}(\nabla F_1)^2(\nabla F_2)^2+\frac{3}{4}(\nabla F_1)^4+\frac{7}{4}(\nabla F_2)^4\\
&\qquad-{2F_1}(\partial^i\partial^jF_1)\big[2(\partial^iF_2)(\partial_jF_2)+(\partial_iF_1)(\partial_jF_1)-F_1(\partial_i\partial_jF_1)\big]\\
&\qquad-{2}{F_1}(\partial_i\partial_jF_2)\big[F_1(\partial^i\partial^jF_2)-3(\partial^iF_2)(\partial^jF_1)\big]
\bigg]
\end{aligned}    
\end{equation}
The right hand side vanishes if $F_1=F_2=F_3$, and the entire equation reduces to the Laplace equation for $F_1^{-1}$.  

The equation of motion with respect to $F_2$ is given by,
\bea\label{CNMeqnF2}
&&\frac{1}{F_1^2F_3^2}\big[F_1F_3(\partial_k\partial^kF_2)-{F_3}(\partial_kF_1)(\partial^kF_3)-{F_1}(\partial_kF_3)(\partial^kF_2)\big]\nn
&&=-\frac{\alpha^\prime}{4}\big[-\frac{20}{F_1^5}(\nabla F_1)^4+
\frac{1}{F_1^2}(\nabla^4 F_1)+\frac{22}{F_1^4}(\partial_l\partial_kF_1)(\partial^kF_1)(\partial^lF_1)\nn
&&\quad+\frac{19}{2F_1^4}(\nabla F_1)^2(\nabla^2 F_1)-\frac{3}{2F_1^3}(\nabla^2 F_1)^2-\frac{4}{F_1^3}(\partial_k\partial_lF_1)(\partial^k\partial^lF_1)\\
&&\quad-\frac{15}{2F_1^3}(\partial_lF_1)(\partial^l\nabla^2 F_1)\big]\nonumber
\eea
and the equation of motion for $F_1$ is 
\bea\label{CNMeqnF1}
&&\frac{1}{2F_1^2F_3^3}\Big[{F_3^2}(\nabla F_1)^2-6{(F_1F_3)^2}(\nabla F_2)^2
(\nabla F_3)^2+{4}{F_3F_1^2}(\nabla^2 F_3)\nn
&&\qquad+2{F_3^2}(\partial_kF_1)(\partial^kF_3)-2{F_1F_3}(\nabla^2 F_1)\Big]\nn
&&={\alpha^\prime}\big[-\frac{20}{F_1^5}(\nabla F_1)^4+
\frac{1}{F_1^2}(\nabla^4 F_1)+\frac{22}{F_1^4}(\partial_l\partial_kF_1)(\partial^kF_1)(\partial^lF_1)\\
&&\qquad+\frac{19}{2F_1^4}(\nabla F_1)^2(\nabla^2 F_1)-\frac{3}{2F_1^3}(\nabla^2F_1)^2-\frac{4}{F_1^3}(\partial_k\partial_lF_1)(\partial^k\partial^lF_1)\nn
&&\qquad-\frac{15}{2F_1^3}(\partial_lF_1)(\partial^l\nabla^2 F_1)\big] \nonumber
\eea
Equations (\ref{CNMeqnF2})--(\ref{CNMeqnF1}) are satisfied for a harmonic function $F_1=F_2=F_3=H^{-1}$, where $H$ is a harmonic function.
\newpage


\begin{thebibliography}{100}

\bibitem{PolchW1} 
J.~Dai, R.~G.~Leigh and J.~Polchinski,
``New Connections Between String Theories,''
Mod. Phys. Lett. A \textbf{4}, 2073-2083 (1989).
\bibitem{PolchW2} 
R.~G.~Leigh,
``Dirac-Born-Infeld Action from Dirichlet Sigma Model,''
Mod. Phys. Lett. A \textbf{4}, 2767 (1989).
\bibitem{PolchW3} 
J.~Polchinski,
``Dirichlet Branes and Ramond-Ramond charges,''
Phys. Rev. Lett. \textbf{75}, 4724-4727 (1995),
arXiv:hep-th/9510017.
\bibitem{PolchW4} 
P.~Horava,
``Strings on World Sheet Orbifolds,''
Nucl. Phys. B \textbf{327}, 461-484 (1989).
\bibitem{PolchW5} 
P.~Horava,
``Background Duality of Open String Models,''
Phys. Lett. B \textbf{231}, 251-257 (1989).
\bibitem{MtheoryW1}
E.~Witten,
``String theory dynamics in various dimensions,''
Nucl. Phys. B \textbf{443}, 85-126 (1995),
arXiv:hep-th/9503124.
\bibitem{MtheoryW2}
%
P.~Horava and E.~Witten,
``Heterotic and type I string dynamics from eleven-dimensions,''
Nucl. Phys. B \textbf{460}, 506-524 (1996),
hep-th/9510209.
%
\bibitem{MtheoryW3}
P.~Horava and E.~Witten,
``Eleven-dimensional supergravity on a manifold with boundary,''
Nucl. Phys. B \textbf{475}, 94-114 (1996)
arXiv:hep-th/9603142.
%
%
\bibitem{BHtdualW1} 
A.~Strominger and C.~Vafa,
``Microscopic origin of the Bekenstein-Hawking entropy,''
Phys. Lett. B \textbf{379}, 99-104 (1996),
arXiv:hep-th/9601029.
\bibitem{BHtdualW2} 
C.~G.~Callan and J.~M.~Maldacena,
  ``D-brane Approach to Black Hole Quantum Mechanics,''
  Nucl.\ Phys.\  B {\bf 472}, 591 (1996), arXiv:hep-th/9602043.
\bibitem{BHtdualW3} 
 G.~T.~Horowitz and A.~Strominger,
  ``Counting States of Near-Extremal Black Holes,''
  Phys.\ Rev.\ Lett.\  {\bf 77}, 2368 (1996), arXiv:hep-th/9602051.
\bibitem{BHtdualW4} 
 J.~C.~Breckenridge, D.~A.~Lowe, R.~C.~Myers, A.~W.~Peet, A.~Strominger and C.~Vafa,
  ``Macroscopic and Microscopic Entropy of Near-Extremal Spinning Black Holes,''
  Phys.\ Lett.\  B {\bf 381}, 423 (1996), arXiv:hep-th/9603078.
\bibitem{AdSCFTw1}
J.~M.~Maldacena,
  ``The large N limit of superconformal field theories and supergravity,''
  Adv.\ Theor.\ Math.\ Phys.\  {\bf 2}, 231 (1998)
  [Int.\ J.\ Theor.\ Phys.\  {\bf 38}, 1113 (1999)], arXiv:hep-th/9711200.
\bibitem{AdSCFTw2}
  S.~S.~Gubser, I.~R.~Klebanov and A.~M.~Polyakov,
  ``Gauge theory correlators from non-critical string theory,''
  Phys.\ Lett.\  B {\bf 428}, 105 (1998), arXiv:hep-th/9802109.
\bibitem{AdSCFTw3}
  E.~Witten,
  ``Anti-de Sitter space and holography,''
  Adv.\ Theor.\ Math.\ Phys.\  {\bf 2}, 253 (1998), arXiv:hep-th/9802150.
\bibitem{AdSCFTw4}
 O.~Aharony, S.~S.~Gubser, J.~M.~Maldacena, H.~Ooguri and Y.~Oz,
  ``Large N field theories, string theory and gravity,''
  Phys.\ Rept.\  {\bf 323}, 183 (2000), arXiv:hep-th/9905111.
  %
\bibitem{TdualW1}
B.~E.~Fridling and A.~Jevicki,
``Dual Representations and Ultraviolet Divergences in Nonlinear $\sigma$ Models,''
Phys. Lett. B \textbf{134}, 70-74 (1984).
\bibitem{TdualW2}
E.~S.~Fradkin and A.~A.~Tseytlin,
``Quantum Equivalence of Dual Field Theories,''
Annals Phys. \textbf{162}, 31 (1985).

\bibitem{NATDw1}
X.~C.~de la Ossa and F.~Quevedo,
``Duality symmetries from non-Abelian isometries in string theory,''
Nucl. Phys. B \textbf{403}, 377-394 (1993),  {arXiv:hep-th/9210021}.
%
\bibitem{NATDw2}
%
E.~Alvarez, L.~Alvarez-Gaume, J.~L.~F.~Barbon and Y.~Lozano,
``Some global aspects of duality in string theory,''
Nucl. Phys. B \textbf{415}, 71-100 (1994),
arXiv:hep-th/9309039.
\bibitem{NATDw3}
%
E.~Alvarez, L.~Alvarez-Gaume and Y.~Lozano,
``On nonAbelian duality,''
Nucl. Phys. B \textbf{424}, 155-183 (1994),
arXiv:hep-th/9403155.
\bibitem{NATDw4}
%
C.~Klimcik and P.~Severa,
``Dual nonAbelian duality and the Drinfeld double,''
Phys. Lett. B \textbf{351}, 455-462 (1995)
arXiv:hep-th/9502122 [hep-th].
%
\bibitem{NATDw5}
C.~Klimcik and P.~Severa,
``Poisson-Lie T duality and loop groups of Drinfeld doubles,''
Phys. Lett. B \textbf{372}, 65-71 (1996)
arXiv:hep-th/9512040 [hep-th].
\bibitem{NATDw6}
%
J.~Balog, P.~Forgacs, N.~Mohammedi, L.~Palla and J.~Schnittger,
``On quantum T duality in sigma models,''
Nucl. Phys. B \textbf{535}, 461-482 (1998)
arXiv:hep-th/9806068 [hep-th].
%
%
\bibitem{SenCvetW1} 
A.~Sen,
  ``Strong - weak coupling duality in four-dimensional string theory,''
  Int.\ J.\ Mod.\ Phys.\ A {\bf 9}, 3707 (1994)
  hep-th/9402002.
\bibitem{SenCvetW2} 
  A.~Sen,
  ``Black hole solutions in heterotic string theory on a torus,''
  Nucl.\ Phys.\ B {\bf 440}, 421 (1995)
  hep-th/9411187.
\bibitem{SenCvetW3} 
A.~Sen,
  ``Duality symmetry group of two-dimensional heterotic string theory,''
  Nucl.\ Phys.\ B {\bf 447}, 62 (1995)
  hep-th/9503057.
\bibitem{SenCvetW4} 
M.~Cvetic and D.~Youm,
  ``Dyonic BPS saturated black holes of heterotic string on a six torus,''
  Phys.\ Rev.\ D {\bf 53}, 584 (1996)
  hep-th/9507090.
\bibitem{SenCvetW5} 
M.~Cvetic and D.~Youm,
  ``All the static spherically symmetric black holes of heterotic string on a six torus,''
  Nucl.\ Phys.\ B {\bf 472}, 249 (1996)
  hep-th/9512127.

\bibitem{TsTQFTw1}
A.~Hashimoto and N.~Itzhaki,
``Noncommutative Yang-Mills and the AdS / CFT correspondence,''
Phys. Lett. B \textbf{465}, 142-147 (1999),
hep-th/9907166.
\bibitem{TsTQFTw2}
J.~M.~Maldacena and J.~G.~Russo,
``Large N limit of noncommutative gauge theories,''
JHEP \textbf{09}, 025 (1999),
hep-th/9908134.
\bibitem{TsTQFTw3}
O.~Lunin and J.~M.~Maldacena,
``Deforming field theories with U(1) x U(1) global symmetry and their gravity duals,''
JHEP \textbf{05}, 033 (2005), {arXiv:hep-th/0502086 [hep-th]}.
\bibitem{TsTQFTw4}
%
S.~Frolov,
``Lax pair for strings in Lunin-Maldacena background,''
JHEP \textbf{05}, 069 (2005), {arXiv:hep-th/0503201 [hep-th]}.
%
\bibitem{TsTQFTw5}
L.~Cornalba, M.~S.~Costa and R.~Schiappa,
``D-brane dynamics in constant Ramond-Ramond potentials, S duality and noncommutative geometry,''
Adv. Theor. Math. Phys. \textbf{9}, no.3, 355-406 (2005), {arXiv:hep-th/0209164 [hep-th]}.
\bibitem{TsTQFTw6}
S.~He and H.~Shu,
``T-duality to Scattering Amplitude and Wilson Loop in Non-commutative Super Yang-Mills Theory,''
JHEP \textbf{08}, 172 (2018), {arXiv:1806.02707 [hep-th]}.


\bibitem{TsTdef}
S.~A.~Frolov, R.~Roiban and A.~A.~Tseytlin,
``Gauge-string duality for superconformal deformations of N=4 super Yang-Mills theory,''
JHEP \textbf{07}, 045 (2005), {arXiv:hep-th/0503192 [hep-th]}.
\bibitem{BetaDefW1}
E.~Imeroni,
``On deformed gauge theories and their string/M-theory duals,''
JHEP \textbf{10}, 026 (2008), {arXiv:0808.1271 [hep-th]}.
\bibitem{BetaDefW2}
S.~J.~van Tongeren,
``On classical Yang-Baxter based deformations of the AdS$_{5}\times$ S$^{5}$ superstring,''
JHEP \textbf{06}, 048 (2015), {arXiv:1504.05516 [hep-th]}.
\bibitem{BetaDefW3}
A.~Sfondrini and S.~J.~van Tongeren,
``$T\bar{T}$ deformations as $TsT$ transformations,''
Phys. Rev. D \textbf{101}, no.6, 066022 (2020), {arXiv:1908.09299 [hep-th]}.
\bibitem{BetaDefW4}
C.~D.~A.~Blair,
``Non-relativistic duality and $T \bar T$ deformations,''
JHEP \textbf{07}, 069 (2020), {arXiv:2002.12413 [hep-th]}.
\bibitem{BetaDefW5}
R.~Borsato, S.~Driezen, J.~M.~Nieto Garc\'\i{}a and L.~Wyss,
``Semiclassical spectrum of a Jordanian deformation of AdS5$\times$S5,''
Phys. Rev. D \textbf{106}, no.6, 066015 (2022), {arXiv:2207.14748 [hep-th]}.



\bibitem{TseytlDFT}
A.~A.~Tseytlin,
``Duality Symmetric Formulation of String World Sheet Dynamics,''
Phys. Lett. B \textbf{242}, 163-174 (1990).
  
\bibitem{DFTw1}
W.~Siegel,
``Superspace duality in low-energy superstrings,''
Phys. Rev. D \textbf{48}, 2826-2837 (1993)
arXiv:hep-th/9305073 [hep-th].
%
\bibitem{DFTw2}
W.~Siegel,
``Two vierbein formalism for string inspired axionic gravity,''
Phys. Rev. D \textbf{47}, 5453-5459 (1993)
arXiv:hep-th/9302036 [hep-th].
\bibitem{DFTw3}
%
C.~Hull and B.~Zwiebach,
``Double Field Theory,''
JHEP \textbf{09}, 099 (2009)
arXiv:0904.4664 [hep-th].
%
\bibitem{DFTw4}
C.~Hull and B.~Zwiebach,
``The Gauge algebra of double field theory and Courant brackets,''
JHEP \textbf{09}, 090 (2009)
arXiv:0908.1792 [hep-th].
\bibitem{DFTw5}
%
O.~Hohm, C.~Hull and B.~Zwiebach,
``Background independent action for double field theory,''
JHEP \textbf{07}, 016 (2010)
arXiv:1003.5027 [hep-th].
\bibitem{DFTw6}
%
O.~Hohm, C.~Hull and B.~Zwiebach,
``Generalized metric formulation of double field theory,''
JHEP \textbf{08}, 008 (2010)
arXiv:1006.4823 [hep-th].
\bibitem{DFTw7}
%
O.~Hohm and S.~K.~Kwak,
``Frame-like Geometry of Double Field Theory,''
J. Phys. A \textbf{44}, 085404 (2011)
arXiv:1011.4101 [hep-th].

\bibitem{DFTclassW1}
M.~J.~Duff,
``Duality Rotations in String Theory,''
Nucl. Phys. B \textbf{335}, 610 (1990).
\bibitem{DFTclassW2}
M.~J.~Duff and J.~X.~Lu,
``Duality Rotations in Membrane Theory,''
Nucl. Phys. B \textbf{347}, 394-419 (1990).
\bibitem{DFTclassW3}
A.~Sen,
``O(d) x O(d) symmetry of the space of cosmological solutions in string theory, scale factor duality and two-dimensional black holes,''
Phys. Lett. B \textbf{271}, 295-300 (1991).
\bibitem{DFTclassW4}
E.~Bergshoeff, B.~Janssen and T.~Ortin,
``Solution generating transformations and the string effective action,''
Class. Quant. Grav. \textbf{13}, 321-343 (1996)
arXiv:hep-th/9506156 [hep-th].
\bibitem{DFTclassW5}
O.~Hohm and S.~K.~Kwak,
``Double Field Theory Formulation of Heterotic Strings,''
JHEP \textbf{06}, 096 (2011)
arXiv:1103.2136 [hep-th].
\bibitem{DFTclassW6}
D.~Geissbuhler, D.~Marques, C.~Nunez and V.~Penas,
``Exploring Double Field Theory,''
JHEP \textbf{06}, 101 (2013)
arXiv:1304.1472 [hep-th].
\bibitem{DFTclassW7}
S.~Demulder, F.~Hassler and D.~C.~Thompson,
``An invitation to Poisson-Lie T-duality in Double Field Theory and its applications,''
PoS \textbf{CORFU2018}, 113 (2019)
arXiv:1904.09992 [hep-th].
\bibitem{DFTclassW8}
D.~C.~Thompson,
``An Introduction to Generalised Dualities and their Applications to Holography and Integrability,''
PoS \textbf{CORFU2018}, 099 (2019)
arXiv:1904.11561 [hep-th].
\bibitem{DFTclassW9}
%
R.~Borsato and S.~Driezen,
``Supergravity solution-generating techniques and canonical transformations of $\sigma$-models from $O(D,D)$,''
JHEP \textbf{05}, 180 (2021),
{arXiv:2102.04498 [hep-th]}.
\bibitem{DFTclassW10}
R.~Borsato, S.~Driezen and F.~Hassler,
``An algebraic classification of solution generating techniques,''
Phys. Lett. B \textbf{823}, 136771 (2021)
arXiv:2109.06185 [hep-th].
\bibitem{DFTclassW11}
F.~Hassler,
``Poisson-Lie T-duality in Double Field Theory,''
Phys. Lett. B \textbf{807}, 135455 (2020)
arXiv:1707.08624 [hep-th].
\bibitem{DFTclassW12}
D.~L\"ust and D.~Osten,``Generalised fluxes, Yang-Baxter deformations and the O(d,d) structure of non-abelian T-duality,''JHEP \textbf{05}, 165 (2018),
arXiv:1803.03971.
\bibitem{DFTclassW13}
A.~Catal-Ozer,
``Non-Abelian T-duality as a Transformation in Double Field Theory,''
JHEP \textbf{08}, 115 (2019), {arXiv:1904.00362 [hep-th]}.

\bibitem{DFTalpW1}
O.~Hohm and B.~Zwiebach,
``On the Riemann Tensor in Double Field Theory,''
JHEP \textbf{05}, 126 (2012)
arXiv:1112.5296 [hep-th].
%
\bibitem{DFTalpW2}
O.~Hohm, W.~Siegel and B.~Zwiebach,
``Doubled $\alpha'$-geometry,''
JHEP \textbf{02}, 065 (2014)
arXiv:1306.2970 [hep-th].
%
\bibitem{DFTalpW3}
O.~Hohm and B.~Zwiebach,
``Double field theory at order $\alpha'$,''
JHEP \textbf{11}, 075 (2014)
arXiv:1407.3803 [hep-th].
%
\bibitem{DFTalpW4}
O.~Hohm and B.~Zwiebach,
``Green-Schwarz mechanism and $\alpha'$-deformed Courant brackets,''
JHEP \textbf{01}, 012 (2015)
arXiv:1407.0708 [hep-th].
\bibitem{DFTalpW5}
O.~A.~Bedoya, D.~Marques and C.~Nunez,
``Heterotic $\alpha$'-corrections in Double Field Theory,''
JHEP \textbf{12}, 074 (2014)
arXiv:1407.0365 [hep-th].
\bibitem{DFTalpW6}
%
A.~Coimbra, R.~Minasian, H.~Triendl and D.~Waldram,
``Generalised geometry for string corrections,''
JHEP \textbf{11}, 160 (2014)
arXiv:1407.7542 [hep-th].
%
\bibitem{DFTalpW7}
K.~Lee,
``Quadratic \ensuremath{\alpha}'-corrections to heterotic double field theory,''
Nucl. Phys. B \textbf{899}, 594-616 (2015)
arXiv:1504.00149 [hep-th].

\bibitem{1507}
D.~Marques and C.~A.~Nunez,
``T-duality and \ensuremath{\alpha}'-corrections,''
JHEP \textbf{10}, 084 (2015)
arXiv:1507.00652 [hep-th].

\bibitem{OGBuscherW1}
T.~H.~Buscher,
``A Symmetry of the String Background Field Equations,''
Phys. Lett. B \textbf{194}, 59-62 (1987).
\bibitem{OGBuscherW2}
T.~H.~Buscher,
``Path Integral Derivation of Quantum Duality in Nonlinear Sigma Models,''
Phys. Lett. B \textbf{201}, 466-472 (1988).


\bibitem{NATD101}
G.~Itsios, C.~Nunez, K.~Sfetsos and D.~C.~Thompson,
``Non-Abelian T-duality and the AdS/CFT correspondence:new N=1 backgrounds,''
Nucl. Phys. B \textbf{873}, 1-64 (2013)
arXiv:1301.6755 [hep-th].

\bibitem{NATDAdSCFTw1}
G.~Itsios, C.~Nunez, K.~Sfetsos and D.~C.~Thompson,
``On Non-Abelian T-Duality and new N=1 backgrounds,''
Phys. Lett. B \textbf{721}, 342-346 (2013)
arXiv:1212.4840 [hep-th].
%
\bibitem{NATDAdSCFTw2}
%
N.~T.~Macpherson,
``Non-Abelian T-duality, $G_2$-structure rotation and holographic duals of $N=1$ Chern-Simons theories,''
JHEP \textbf{11}, 137 (2013)
arXiv:1310.1609 [hep-th].
%
\bibitem{NATDAdSCFTw3}
K.~Sfetsos and D.~C.~Thompson,
``New ${\cal N} = 1$ supersymmetric $AdS_5$ backgrounds in Type IIA supergravity,''
JHEP \textbf{11}, 006 (2014)
arXiv:1408.6545 [hep-th].
%
\bibitem{NATDAdSCFTw4}
\"O.~Kelekci, Y.~Lozano, N.~T.~Macpherson and E.~\'O.~Colg\'ain,
``Supersymmetry and non-Abelian T-duality in type II supergravity,''
Class. Quant. Grav. \textbf{32}, no.3, 035014 (2015)
arXiv:1409.7406 [hep-th].
%
\bibitem{NATDAdSCFTw5}
N.~T.~Macpherson, C.~N\'u\~nez, L.~A.~Pando Zayas, V.~G.~J.~Rodgers and C.~A.~Whiting,
``Type IIB supergravity solutions with AdS$_{5}$ from Abelian and non-Abelian T dualities,''
JHEP \textbf{02}, 040 (2015)
arXiv:1410.2650 [hep-th].
%
\bibitem{NATDAdSCFTw6}
%
H.~Dimov, S.~Mladenov, R.~C.~Rashkov and T.~Vetsov,
``Non-abelian T-duality of Pilch-Warner background,''
Fortsch. Phys. \textbf{64}, 657-673 (2016)
arXiv:1511.00269 [hep-th].
%
\bibitem{NATDAdSCFTw7}
B.~Hoare and A.~A.~Tseytlin,
``Homogeneous Yang-Baxter deformations as non-abelian duals of the $\mathrm{AdS}_5 \sigma$-model,''
J. Phys. A \textbf{49}, no.49, 494001 (2016)
arXiv:1609.02550 [hep-th].
%
\bibitem{NATDAdSCFTw8}
%
R.~Borsato and L.~Wulff,
``On non-abelian T-duality and deformations of supercoset string sigma-models,''
JHEP \textbf{10}, 024 (2017)
arXiv:1706.10169 [hep-th].
%
\bibitem{NATDAdSCFTw9}
R.~Borsato and L.~Wulff,
``Non-abelian T-duality and Yang-Baxter deformations of Green-Schwarz strings,''
JHEP \textbf{08}, 027 (2018)
arXiv:1806.04083 [hep-th].
%
\bibitem{NATDAdSCFTw10}
G.~Itsios, J.~M.~Pen\'\i{}n and S.~Zacar\'\i{}as,
``Spin-2 excitations in Gaiotto-Maldacena solutions,''
JHEP \textbf{10}, 231 (2019)
arXiv:1903.11613 [hep-th].
%
\bibitem{NATDAdSCFTw11}
B.~Hoare and A.~A.~Tseytlin,
``Homogeneous Yang-Baxter deformations as non-abelian duals of the ${AdS}_5 \sigma$-model,''
J. Phys. A \textbf{49}, 494001 (2016),
{arXiv:1609.02550 [hep-th]}.
%
\bibitem{NATDAdSCFTw12}
N.~T.~Macpherson, C.~N\'u\~nez, L.~A.~Pando Zayas, V.~G.~J.~Rodgers and C.~A.~Whiting,
``Type IIB supergravity solutions with AdS$_{5}$ from Abelian and non-Abelian T dualities,''
JHEP \textbf{02}, 040 (2015),
{arXiv:1410.2650 [hep-th]}.

\bibitem{BdR-1w1}
E.~Bergshoeff and M.~de Roo,
``Supersymmetric Chern-simons Terms in Ten-dimensions,''
Phys. Lett. B \textbf{218}, 210-215 (1989).
%
\bibitem{BdR-1w2}
E.~A.~Bergshoeff and M.~de Roo,
``The Quartic Effective Action of the Heterotic String and Supersymmetry,''
Nucl. Phys. B \textbf{328}, 439-468 (1989)

\bibitem{MT}
R.~R.~Metsaev and A.~A.~Tseytlin,
``Order alpha-prime (Two Loop) Equivalence of the String Equations of Motion and the Sigma Model Weyl Invariance Conditions: Dependence on the Dilaton and the Antisymmetric Tensor,''
Nucl. Phys. B \textbf{293}, 385-419 (1987)



\bibitem{BdR=MT}
W.~A.~Chemissany, M.~de Roo and S.~Panda,
``alpha'-Corrections to Heterotic Superstring Effective Action Revisited,''
JHEP \textbf{08}, 037 (2007)
arXiv:0706.3636 [hep-th].

\bibitem{LuShah}
O.~Lunin and P.~Shah,
``Geometries with twisted spheres and non-abelian T-dualities,''
JHEP \textbf{03}, 102 (2024),
arXiv:2311.14285 [hep-th].

\bibitem{TsTold}
J.~G.~Russo and A.~A.~Tseytlin,
``Exactly solvable string models of curved space-time backgrounds,''
Nucl. Phys. B \textbf{449}, 91-145 (1995),
arXiv:hep-th/9502038;\\
J.~G.~Russo and A.~A.~Tseytlin,
``Magnetic flux tube models in superstring theory,''
Nucl. Phys. B \textbf{461}, 131-154 (1996),
arXiv:hep-th/9508068 [hep-th].

\bibitem{2007.07902}
R.~Borsato and L.~Wulff,
``Quantum Correction to Generalized $T$ Dualities,''
Phys. Rev. Lett. \textbf{125}, no.20, 201603 (2020)
arXiv:2007.07902 [hep-th].

\bibitem{2007.07897}
F.~Hassler and T.~Rochais,
``$\alpha'$-Corrected Poisson-Lie T-Duality,''
Fortsch. Phys. \textbf{68}, no.9, 2000063 (2020)
arXiv:2007.07897 [hep-th].

\bibitem{2007.09494}
T.~Codina and D.~Marques,
``Generalized Dualities and Higher Derivatives,''
JHEP \textbf{10}, 002 (2020)
arXiv:2007.09494 [hep-th].

\bibitem{fundRiem}
F.~Hassler and Y.~Sakatani,
``Hierarchy of curvatures in exceptional geometry,''
Phys. Rev. D \textbf{109}, no.10, 106002 (2024),
arXiv:2311.12095 [hep-th].

\bibitem{Cartan1}
E.~Cartan and J.A.~Schouten, ``On the Geometry of the group manifold of simple and semisimple groups,'' Proc. Amsterdam 29 (1926), 803--815.
\bibitem{Cartan2}
E.~Cartan and J.A.~Schouten,  ``On Riemannian manifolds admitting an absolute parallelism,'' Proc. Amsterdam 29 (1926), 933--946.

\bibitem{CHS}
C.~G.~Callan, Jr., J.~A.~Harvey and A.~Strominger,
`World sheet approach to heterotic instantons and solitons,''
Nucl. Phys. B \textbf{359}, 611-634 (1991)

\bibitem{2206.10640}
S.~Hronek, L.~Wulff and S.~Zacarias,
``The \ensuremath{\alpha}'$^{2}$ correction from double field theory,''
JHEP \textbf{11}, 090 (2022)
arXiv:2206.10640 [hep-th]

\bibitem{AlpObstr}
S.~Hronek and L.~Wulff,
``$O(D,D)$ and the string $\alpha'$ expansion: an obstruction,''
JHEP \textbf{04}, 013 (2021)
arXiv:2012.13410 [hep-th];\\
L.~Wulff,
``Tree-level $R^4$ correction from $O(d,d)$: NS-NS five-point terms,''
arXiv:2406.15240 [hep-th].

\bibitem{CNMw1}
G.~T.~Horowitz and A.~A.~Tseytlin,
``On exact solutions and singularities in string theory,''
Phys. Rev. D \textbf{50}, 5204-5224 (1994)
arXiv:hep-th/9406067 [hep-th].
%
\bibitem{CNMw2}
G.~T.~Horowitz and A.~A.~Tseytlin,
``A New class of exact solutions in string theory,''
Phys. Rev. D \textbf{51}, 2896-2917 (1995)
arXiv:hep-th/9409021 [hep-th].

\bibitem{Viser16}
P.~Bueno, P.~A.~Cano, V.~S.~Min and M.~R.~Visser,
``Aspects of general higher-order gravities,''
Phys. Rev. D \textbf{95}, no.4, 044010 (2017),
arXiv:1610.08519 [hep-th].

\bibitem{OldNATD1w1}
T.~Curtright and C.~K.~Zachos,
``Currents, charges, and canonical structure of pseudodual chiral models,''
Phys. Rev. D \textbf{49}, 5408-5421 (1994)
arXiv:hep-th/9401006 [hep-th].
%
\bibitem{OldNATD1w2}
Y.~Lozano,
``NonAbelian duality and canonical transformations,''
Phys. Lett. B \textbf{355}, 165-170 (1995)
arXiv:hep-th/9503045 [hep-th].
\bibitem{OldNATD1w3}
%
Y.~Lozano,
``Duality and canonical transformations,''
Mod. Phys. Lett. A \textbf{11}, 2893-2914 (1996)
arXiv:hep-th/9610024 [hep-th].
%
\bibitem{OldNATD1w4}
K.~Sfetsos,
``NonAbelian duality, parafermions and supersymmetry,''
Phys. Rev. D \textbf{54}, 1682-1695 (1996)
arXiv:hep-th/9602179 [hep-th].

\bibitem{1012}
K.~Sfetsos and D.~C.~Thompson,
``On non-abelian T-dual geometries with Ramond fluxes,''
Nucl. Phys. B \textbf{846}, 21-42 (2011)
arXiv:1012.1320 [hep-th].
\bibitem{1104w1}
K.~Sfetsos,
``Duality invariant class of two-dimensional field theories,''
Nucl. Phys. B \textbf{561}, 316-340 (1999)
arXiv:hep-th/9904188 [hep-th].
\bibitem{1104w2}
Y.~Lozano, E.~O Colgain, K.~Sfetsos and D.~C.~Thompson,
``Non-abelian T-duality, Ramond Fields and Coset Geometries,''
JHEP \textbf{06}, 106 (2011)
arXiv:1104.5196 [hep-th].
%
\bibitem{1104w3}
K.~Sfetsos,
``Integrable interpolations: From exact CFTs to non-Abelian T-duals,''
Nucl. Phys. B \textbf{880}, 225-246 (2014),
{arXiv:1312.4560 [hep-th]}.

\bibitem{9705193w1}
K.~A.~Meissner,
``Symmetries of higher order string gravity actions,''
Phys. Lett. B \textbf{392}, 298-304 (1997)
arXiv:hep-th/9610131 [hep-th].
%
\bibitem{9705193w2}
N.~Kaloper and K.~A.~Meissner,
``Duality beyond the first loop,''
Phys. Rev. D \textbf{56}, 7940-7953 (1997)
arXiv:hep-th/9705193 [hep-th].
%
\bibitem{9705193w3}
N.~Kaloper and K.~A.~Meissner,
``Tailoring T duality beyond the first loop,''
arXiv:hep-th/9708169 [hep-th].

\bibitem{1608}
S.~Sasaki and M.~Yata,
``Non-geometric Five-branes in Heterotic Supergravity,''
JHEP \textbf{11}, 064 (2016),
arXiv:1608.01436 [hep-th].






\end{thebibliography}
\end{document}